\documentclass[fleqn,10pt]{wlscirep}
\usepackage[utf8]{inputenc}
\usepackage[T1]{fontenc}
\usepackage{subfig}
\usepackage{booktabs}
\usepackage{longtable}
\title{Nordic vaccination strategies face/off via age range comparative analysis on key indicators of COVID-19 severity and healthcare stress level}

\author[1]{Anna Sigridur Islind}
\author[1*]{María \'Oskarsd\'ottir}
\author[2,3]{Corentin Cot}
\author[2,3]{Giacomo Cacciapaglia}
\author[4,5]{Francesco Sannino}
\affil[1]{\mbox{Department of Computer Science, Reykjavík University, Menntavegur 1, 102 Reykjavík, Iceland}}
\affil[2]{\mbox{Institut de Physique des deux Infinis de Lyon (IP2I),  UMR5822, CNRS/IN2P3, F-69622, Villeurbanne, France}}
\affil[3]{\mbox{University of Lyon, Universit{\' e} Claude Bernard Lyon 1,  F-69001, Lyon, France}}
\affil[4]{CP3-Origins \& the Danish Institute for Advanced Study, University of Southern Denmark, Campusvej 55, DK-5230 Odense, Denmark}
\affil[5]{ Dipartimento di Fisica E. Pancini, Universit\`a di Napoli Federico II \& INFN sezione di Napoli, Complesso Universitario di Monte S. Angelo Edificio 6, via Cintia, 80126 Napoli, Italy}

\affil[*]{mariaoskars@ru.is}


\begin{abstract}
In this paper we analyze the impact of vaccinations on spread of the COVID-19 virus for different age groups. More specifically we examine the deployment of vaccines in the Nordic countries in a comparative analysis where we analyze factors such as healthcare stress level and severity of disease through new infections, hospitalizations, intensive care unit (ICU) occupancy and deaths. Moreover, we analyze the impact of the various vaccine types, vaccination rate on the spread of the virus in each age group for Denmark, Finland, Iceland, Norway and Sweden from the start of the vaccination period in December 2020 until the end of August 2021. We perform a three-fold analysis: i) frequency analysis of infections and vaccine rates by age groups, ii) rolling correlations between vaccination strategies, severity of COVID-19 and healthcare stress level and; iii) we also employ the epidemic Renormalization Group (eRG) framework. The latter is used to mathematically model wave structures, as well as the impact of vaccinations on wave dynamics. We further compare the Nordic countries with England. Our main results are the quantification of the impact of the vaccination campaigns on age groups epidemiological data, across countries with high vaccine uptake. The data clearly shows that vaccines markedly reduce the number of new cases and the risk of serious illness. 

\end{abstract}
\begin{document}

\flushbottom
\maketitle
%
%
\thispagestyle{empty}

\section*{Highlights}
\begin{itemize}
\item
We quantify and compare the temporal impact of vaccinations across age groups in Nordic countries where there is a high uptake. 
\item
We analyze severity and healthcare stress levels through data on new infections, vaccinations, hospitalizations, intensive care unit (ICU) occupancy and deaths through dynamic correlations.
\item
We also show that when the vaccination rate in a specific age group has reached a plateau, a drop in the new infection rate follows and decreases drastically for that age group with substantially lower disease severity.
\item
The oldest age group, that was vaccinated early are getting less infected in the first post-vaccine pandemic wave whereas the youngest, and least vaccinated are getting increasingly more infected.
\item
We use the eRG framework to analyze and quantify the epidemiological data. The analysis confirms that the vaccination campaigns were effective in reducing both the infection rate and the total number of infections per age group.
\item
A total vaccination is recommended, for all age groups to minimize the risk of severe COVID-19 illness. 
\end{itemize}

\section*{Introduction}
In December 2019, a severe respiratory disease emerged in Wuhan, China \cite{wu2020new}. Since the first identified case of COVID-19 was reported on 8. December 2019 \cite{huang2020clinical,zhu2020novel}, the virus has spread quickly, causing a worldwide health crisis. The spread has occurred in waves, and the virus evolves and mutates, and consequently we have yet not seen the final numbers of ebb and flow with 215 million confirmed cases and 4.48 million confirmed deaths to date. To halt the spread, various non-pharmaceutical methods were put into place, such as limited travel in combination with social distancing as important measures to slow down the short-term spread \cite{islind2020changes}. Eventually, pharmaceutical interventions were administered through vaccinations, to slow down the long-term spread \cite{cacciapaglia2020you, moore2021vaccination}. The first vaccine doses were administered in December 2020, a year after the pandemic started. Since then, vaccinations have been seen as the utmost important strategy for containing the spread of the virus, for minimizing risks for citizens and ultimately for slowly alleviating the non-pharmaceutical restrictions that have been put in place \cite{moore2021vaccination}. As of August 27 2021, 33 \% of the world population has received at least one dose of a COVID-19 vaccine and 25.1 \% are fully vaccinated.

The virus is particularly vigilant and a paper based on data from the European mortality monitoring activity network (EuroMOMO) concluded that all-cause excess mortality in Europe until April 2020, 90\% of excess mortality was due to COVID-19 in the age group of 64 and older \cite{vestergaard2020excess}, pointing towards the notion that age should outline the primary criteria for vaccination prioritisation. Other papers have showed that super-spreaders are the important link in taming the pandemic, as they spread the virus to the healthy and the ill and they should therefore be vaccinated first \cite{lewis2021superspreading,beldomenico2020superspreaders,giubilini2020covid}. 

As of August 27th 2021, a total of 623 million COVID-19 vaccine doses have been distributed within the European Union and European Economic Area (EU/EEA) and 530 million of those doses have already been given to citizens. In the EU/EEA, the cumulative vaccine uptake since the first doses were administered is 75,9\% for first dose and 67,3\% of citizens are fully vaccinated \footnote{https://vaccinetracker.ecdc.europa.eu/public/extensions/COVID-19/vaccine-tracker.html}. Furthermore, the vaccination uptake is especially high in some European countries. In the Nordic countries (Denmark, Finland, Iceland, Norway and Sweden), the vaccination uptake is notably high. In Iceland 91.2\% of the population has received one dose and 86.5\% is fully vaccinated. In Denmark 88.6\% of the population has received one dose and 83.5\% is fully vaccinated. In Finland 84.2\% of the population has received one dose and 59.3\% is fully vaccinated. In Norway 87 \% of the population has received one dose and 56.6\% is fully vaccinated. In Sweden 81.7\% of the population has received one dose and 66.6\% is fully vaccinated to date. Likewise, England had high numbers early. These statistics are for the population of 18 years old and above. 

Due to limited supply of vaccine, and because the virus is highly contagious, the way the vaccine for COVID-19 is prioritized in each country, is utmost important \cite{bubar2021model}. The strategies for prioritization vary between countries, where some countries administer the vaccine first to those of higher age or with serious health issues while other countries prioritize those in high-exposure occupations which are considered a greater risk for spreading the virus \cite{sultana2020potential,goldstein2021vaccinating}. Furthermore, the time between doses, and whether to focus on administering one dose or fully vaccinating people in each age group also varies, as can be seen from the statistics from the Nordic countries we consider in this paper. Moreover, due to severe side-effects reported, some countries have resorted to a ban of administering certain vaccine types to their citizens while other countries have chosen a path of vaccinating with a wide variety of vaccines (with a mix of the vaccines from the producers Pfizer, Moderna, Astra Zeneca and Johnson\&Johnson), arguing that the effects of COVID-19 are more severe than the vaccine side-effects. 

These extremes in choices related to vaccine brands offered, is especially visible in the Nordic countries, where each country has chosen their own path. Because of that, and due to high vaccine uptake to date in the Nordic countries, they are interesting to examine closer. Based on that, we examine the link between new infections and vaccination uptake through open data in the Nordic countries and ask the following two research questions: i) how are the wave structures effected by the vaccination rate within and between various age groups in countries with high vaccination uptake? ii) how are the key indicators of COVID-19 severity and healthcare stress level correlated?
The aim is to examine the dynamic correlations between infections and vaccination rates within and between age groups in the Nordic countries. More specifically, we factor in new infections and vaccine uptake for each age group and illustrate the differences in the vaccination strategy chosen, and the effects in new cases, the vaccination rate in each age group, seriousness of illness in the population and healthcare stress level. Furthermore, we employ the epidemic Renormalization Group (eRG) framework to illustrate wave structures and the impact of vaccinations on the wave structure and we contrast these findings with data from England \cite{cot2021impact,cacciapaglia2020multiwave}. The main contribution of this paper is the quantification of the effects of the vaccine on age groups, across countries with high vaccine uptake. 


\section*{Methods}

In this section we review our methods, and the data used for this paper. Our data is extracted from open source online repositories on the virus spread through new cases reported, the seriousness of the infection through hospitalizations and deaths, the vaccinations in each country for each age group and vaccine type, and the interplay between these variables. The data analysis performed for this study is three-fold: i) frequency analysis of infections and vaccine rates by age groups, ii) rolling correlations between vaccination strategies, severity of COVID-19 and healthcare stress level and; iii) we also employ the epidemic Renormalization Group (eRG) framework.

\subsection*{Vaccine uptake}
The vaccine uptake in the the Nordic countries has been high, as stated earlier. In the data analyzed in this paper we look at weekly numbers of doses administered for all vaccine types allowed and administered in the Nordic countries. More specifically we look at dosed administered of Vaxzevria – produced by AstraZeneca (AZ), Ad26.COV 2.5 – produced by Janssen/Johnson\&Johnson (JANSS), mRNA-1273 – produced by Moderna (MOD) and Comirnaty – produced by Pfizer/BioNTech (COM) in Denmark, Finland, Iceland and Sweden. The information for which vaccine types are used, is not available for Norway, so it is not included in the table. However, Norway has primarily been using mRNA vaccines, (see table \ref{tab:overview}) 

\begin{table}[ht]
 \caption{Overview of the vaccine types used in the Nordic countries}
    \label{tab:overview}
    \centering
    \begin{tabular}{l|cccccccc}
    \toprule
Country&\multicolumn{4}{c}{Vaccine}&Comment \\
  &COM&MOD&AZ&JANSS&& \\
  \midrule
       Denmark&yes&yes&on hold&on hold&AZ \& JANSS were only used in the beginning\\ 
Finland&yes&limited&limited&no&AZ was limited to age 60+ \\
Iceland&yes&yes&limited&yes&AZ was limited to age 55+ \\
Norway&yes&yes&no&no&Decided to exclusively use mRNA vaccines\\
Sweden&yes&yes&limited&on hold&AZ was limited to age 65+\\
         \bottomrule
    \end{tabular}
   
\end{table}

\subsection*{Data description}
The goal of this research is to study the relationship between the progress of the pandemic in terms of new infections and vaccination rate per age group. As stated earlier, we focus on the Nordic countries, which have different vaccination strategies. We collected data from the website of the European Center of Disease Control (ECDC). ECDC provides data that is open source and is collected through The European Surveillance System (TESSy). EU/EEA Member States are requested to report basic indicators (e.g., number of vaccine doses distributed by manufacturers, number of first and second doses administered) alongside new infections, hospitalizations, ICU occupations and deaths in their country. This data was downloaded and processed in order to conduct a retrospective analysis. For the purpose of this paper we use weekly data, divided by age group including the features i) Number of new infections; ii) vaccinations for the first dose; iii) vaccinations for fully vaccinated; iv) vaccination type for the first dose and; v) vaccination type for fully vaccinated. Moreover, we use weekly data for the entirety of the population (not divided into age groups, since age groups were not available for this data): vi) number of hospitalizations for the population as a whole; vii) ICU occupations and; viii) deaths for the population as a whole. The data utilized for the first part of the analysis is from the date when the vaccination campaign started in each country (end of December 2020) whereas the data used for the eRG part of the analysis is for the entirety of COVID-19 infections in each of the countries (since the first case was reported for each of the countries in this study). The age groups for new infections and vaccinations do not match, as seen in Table \ref{T:ageGroups}. 

\begin{table}[]
    \centering
    \caption{The age groups for new infections and vaccinations datasets. }
    \label{T:ageGroups}
    \begin{tabular}{cccccccc}\toprule
    Data set& \multicolumn{7}{c}{Age Groups}\\\midrule
         New Infections&$<15$yr&$15-24$yr&$25-49$yr&$50-64$yr&$65-79$yr&$80+$yr  \\
         Vaccinations&$<18$yr&$18-24$yr&$25-49$yr&$50-59$yr&$60-69$yr&$70-79$yr&$80+$yr  \\ \bottomrule
    \end{tabular}
\end{table}

\subsection*{Rolling correlations}
The first and second part of the analysis illustrated in this paper is based
on frequency analysis of infections and vaccine rates by age groups on the one hand, and on rolling correlations between vaccination strategies, severity of COVID-19 and healthcare stress level on the other hand. The observation period is from the last week of 2020 until week 35 in 2021. 

Rolling correlations show the correlation between two time series, through a rolling window \cite{zivot2003rolling,polanco2020rolwinmulcor}. One of the documented benefits of such an approach is to visualize the correlation change over time \cite{chin2021correlation}. Due to the fact that the pandemic has been ongoing for a while with significant shifts in the wave structure as the number of new cases increase and decrease, the rolling correlation facilitates the inspection of shifts in trends and signals of events that have occurred causing two correlated time series to deviate from each other. Uptake of vaccines could be such an event. More specifically, if there is a relationship to be found, the rolling correlation allows us to model the changes in that relationship over time, through correlations. Therefore, understanding rolling correlations for this type of data over time is both insightful and beneficial. Because the relationships between the variables included in this paper have changed rapidly, and more importantly, have changed over time, we use rolling correlations to show snapshots of those changes, see Section \ref{sec:Rolling}.

For our analysis, we calculate pairwise correlations between the various weekly time series. Due to the non-static nature of the correlation over time, we consider rolling correlations and calculate the correlation between two time series with four weeks of data, and a moving window of one week. This allows us to detect events and shifts in trends, as the pandemic evolves and vaccine update increases. For this we use the Pearson correlation coefficient \cite{benesty2009pearson}. 

We visualize the rolling correlations using networks, where the variables are the nodes and the correlations are the edges. A positive correlation is denoted with a blue edge and a negative one with a red edge. Furthermore, the edge widths show the magnitude of the correlations, a thicker edge means a higher absolute value. For the sake of readability, we remove all edges that have an absolute value below 0.6. In addition to that, we only look at correlations between the new infections per age group and whole population indicators on the one hand and vaccine numbers per age group and whole population indicators on the other hand. Whole population indicators are weekly values for the total number of new infections ('New Infections'), total number of people who have received the first dose ('Total first dose'), the sum of the number of COVID-19 patients in hospital on a given day per week ('Hospital Occupancy'), the sum of the number of COVID-19 patients in ICU on a given day per week ('ICU Occupancy'), and the weekly count of deaths due to COVID-19 ('Deaths'). 

Finally, note that the data for Iceland and the subsequent analysis of the data shows significantly different behaviour in some of the analysis due to limited data points (the population size is 340k and when the new infection rate is low, there is low significance in the analysis due to scarcity in the data). 

\subsection*{Mathematical modelling}
The second part of the analysis illustrated in this paper is based on the employment of the \emph{epidemic Renormalization Group} (eRG) framework, recently developed in \cite{DellaMorte:2020wlc,Cacciapaglia:2020mjf}. It can be mapped \cite{Cacciapaglia:2020mjf,DellaMorte:2020qry} into a time-dependent compartmental model of the SIR type  \cite{Kermack:1927}. The eRG framework provides a single first order differential equation, apt to describing the time-evolution of the cumulative number of infected cases in an isolated region \cite{DellaMorte:2020wlc}. It has been extended in~\cite{Cacciapaglia:2020mjf,cot2021impact} to include interactions among multiple regions of the world.
The main advantage over SIR models is its simplicity, and the fact that it relies on symmetries of the system instead of a detailed description. As a result, no computer simulation is needed in order to understand the time-evolution of the epidemic even at large scales \cite{Cacciapaglia:2020mjf,cot2021impact}.
Recently, the framework has been extended to include the multi-wave pattern \cite{cacciapaglia2020evidence,cacciapaglia2020multiwave} observed in the COVID-19 and other pandemics \cite{1918influenza} and even more recently to include vaccination impact \cite{cot2021impact,cacciapaglia2020you}.

The Renormalization Group approach \cite{Wilson:1971bg,Wilson:1971dh} has a long history in physics with impact from particle to condensed matter physics and beyond. Its application to epidemic dynamics is complementary to other approaches \cite{LI2019566,ZHAN2018437,PERC20171, WANG20151,WANG20161,Danby85,Brauer2019,Miller2012,Murray,Fishman2014,Pell2018}.

\begin{figure}
    \centering
    \includegraphics[scale=0.20]{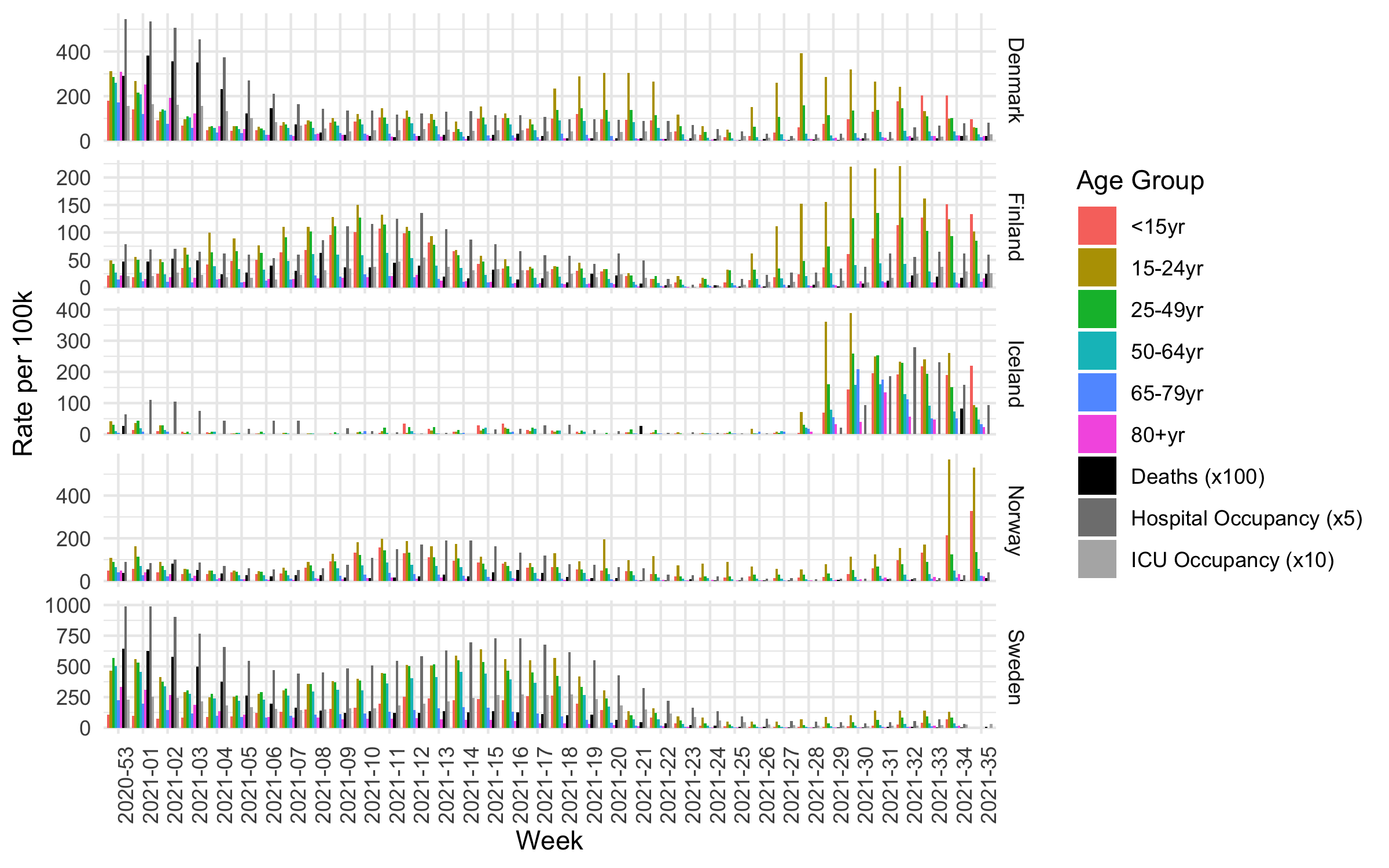}
    \caption{Number of new infections per 100 thousand for each age group, deaths per 10 million, hospital occupancy per 500 thousand and ICU occupancy per million for the five Nordic countries from the end of December 2020 until September 2021.}
    \label{fig:Per100kAgeGroup}
\end{figure}

\section*{Results}

\subsection*{New Infections}
The weekly infection and hospitalization data allows us to view the progression of the pandemic.
Figure \ref{fig:Per100kAgeGroup} shows the weekly numbers of new infections per age group (normalized w.r.t. per 100000 of the age groups' population) during the observation period, as well as the weekly number of deaths (shown x100 in the analysis), ICU occupancy (shown x10 in the analysis) and hospital occupancy (shown x5 in the analysis).

In this figure we can see the waves of the pandemic for the various age groups in the five Nordic countries. In Denmark, Iceland and Sweden, the wave is going down at the start of the observation period and in all countries new waves appear, lead by the younger age groups. This is clear from the high values of the yellow and green bars, which represent people aged 15-24 and 25-49, respectively. However, this effect is less prominent in Sweden. 
What is also clear from these figures is that the number of hospitalizations and deaths are almost non-existent at the end of the observation period, despite the drastic increase in the number of new infections. This is a direct consequence of the high vaccination uptake. We note that this trend is less in Finland, hospitalisations and deaths are still substantial at the end of the period.

\begin{figure}
    \centering
    \includegraphics[scale=0.20]{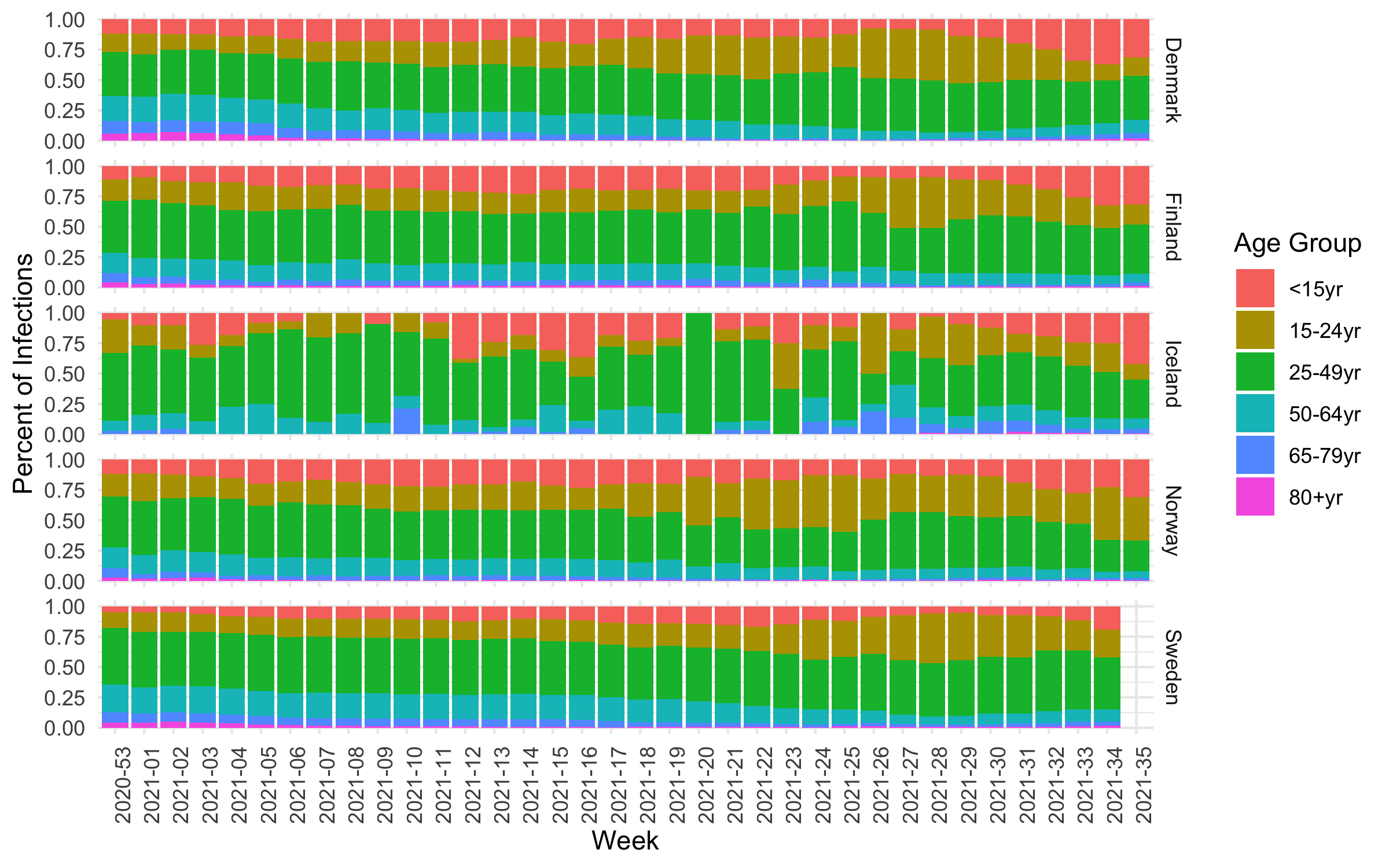}
    \caption{Percentage of each age group of all new infections per week in the five countries.}
    \label{fig:percentAgeGroup}
\end{figure}

Figure \ref{fig:percentAgeGroup} shows the ratio for each age group of the total number of new infection per week. The age group 25-49 usually makes up the largest portion, while we can see some changes in the other age groups. For example, the portion of aged 15-24 year increases over time, whereas the portion of the older generations, aged 65 and older instead diminishes. The impact of the high vaccination uptake in all five countries among the older generations is therefore clear.

\subsection*{Vaccine uptake}
\begin{figure}
    \centering
    \includegraphics[scale=0.2]{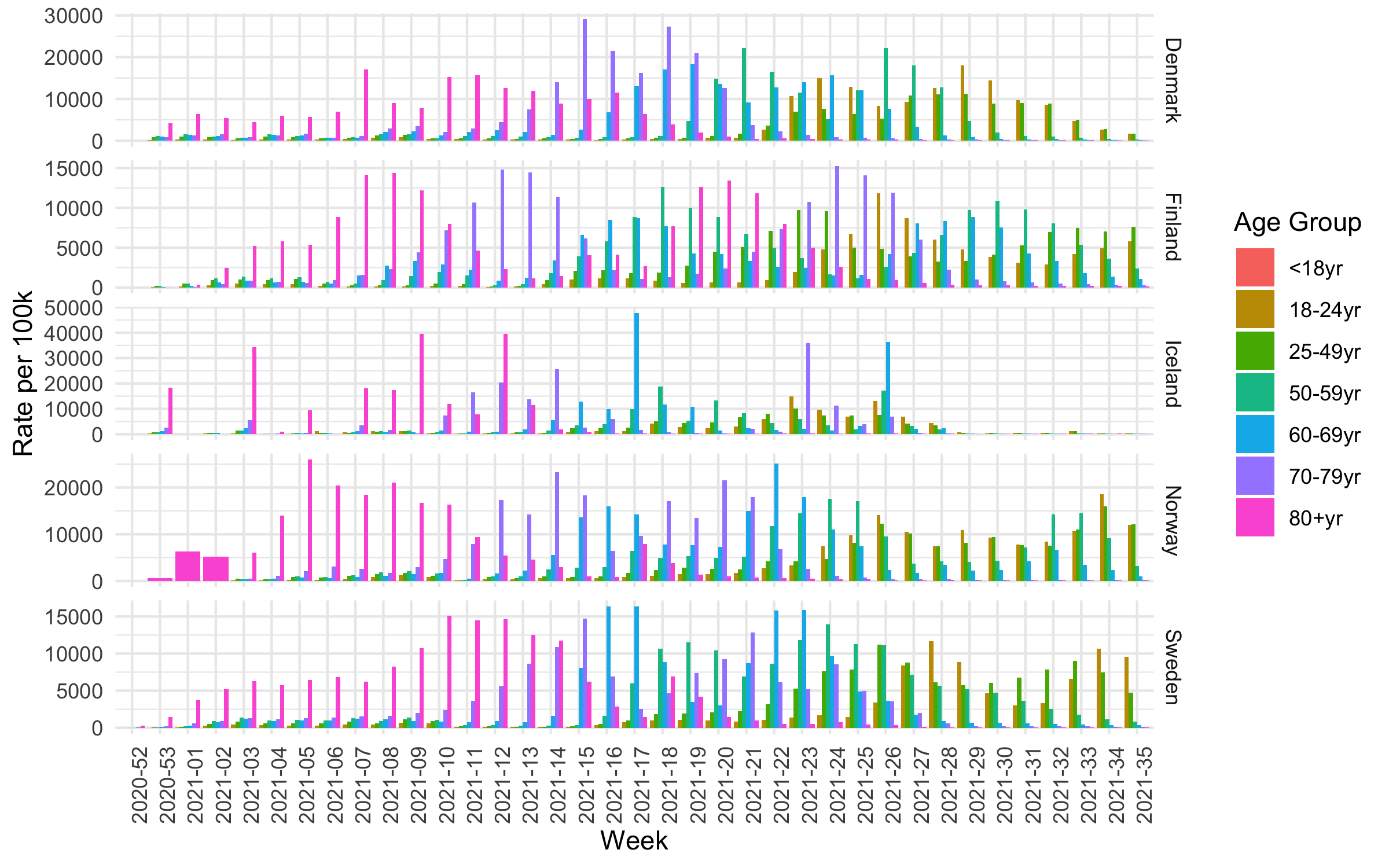}
    \caption{Number of vaccine doses per 100 thousand per age group given each week.}
    \label{fig:vaccineAgeGroup}
\end{figure}

Next, we analyze the vaccination uptake in the five Nordic countries. 
Figure \ref{fig:vaccineAgeGroup} shows the number of vaccines administered per week, divided by age group. Note that these are the numbers for both the first and the second dose.
For each age group, the distribution is roughly bi-modal, which corresponds to the first and second dose of the vaccine.
All countries show a similar trend. First the oldest generation (Age 80+, shown in pink) is vaccinated, followed by people in the other age groups. However, the time delay until younger groups are vaccinated varies between countries. For example, in Finland the 70-79 age group peaks in week 12, whereas in Denmark and Norway this peak happens in week 15. In Denmark, the 18-24 age group is vaccinated in week 22, whereas in Norway there is a delay, and this same age group is vaccinated in week 26.

\begin{figure}
    \centering
    \includegraphics[scale=0.12]{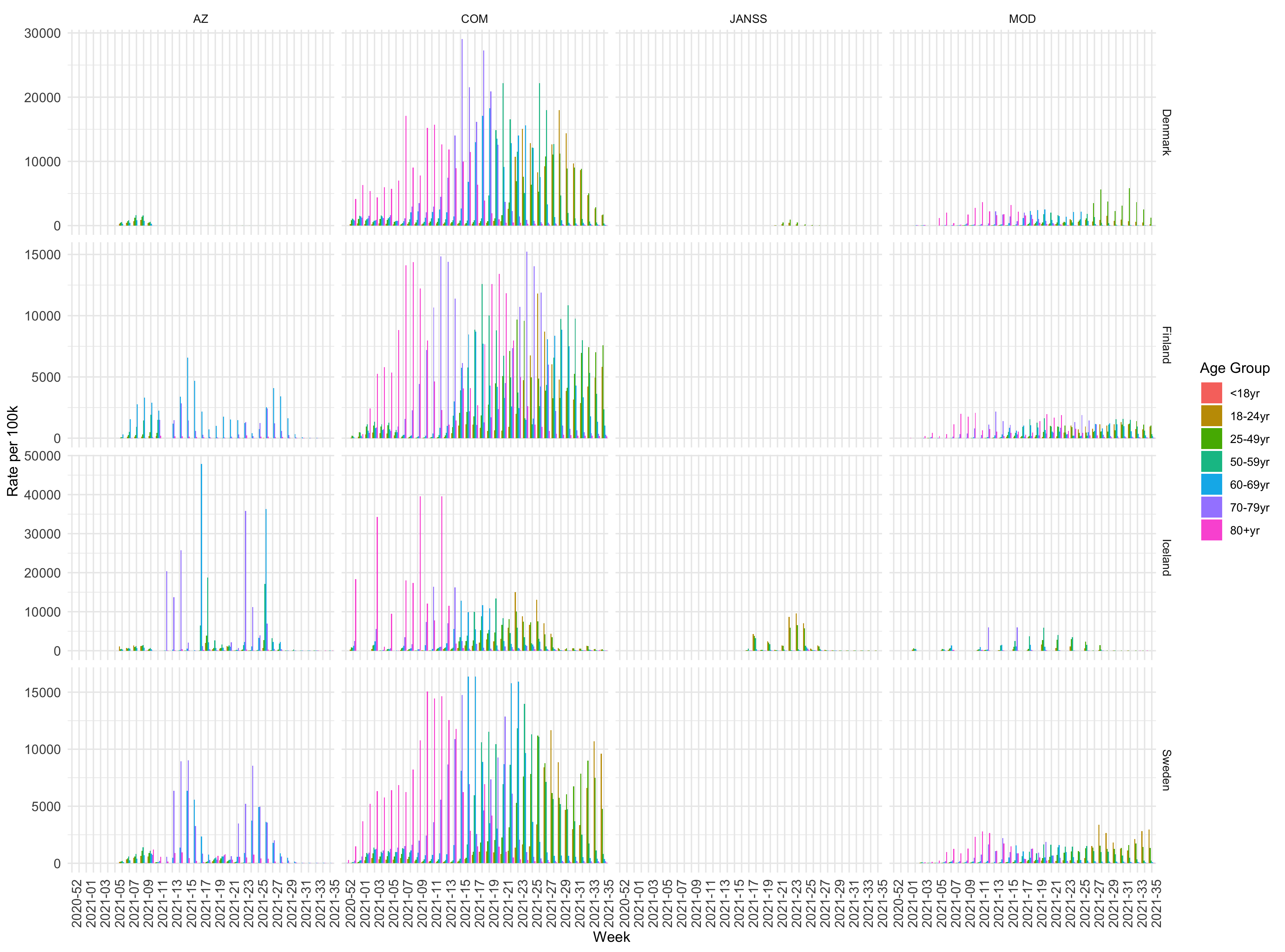}
    \caption{Number of vaccine doses per 100 thousand per age group and vaccine type for each week.}
    \label{fig:vaccineAgeGroupType}
\end{figure}

We take a closer look at the vaccination uptake, by looking at the trends for various vaccine types, see Figure \ref{fig:vaccineAgeGroupType}.
To do that we use data which outlines weekly number of doses administered of different vaccine types (AZ, JANSS, COM and MOD) in Denmark, Finland, Iceland and Sweden.
This information is not available for Norway, so it is not included in the figure. The numbers in this figure are shown per 100 thousand. Across age groups and countries the figure shows that COM is the most commonly used vaccine and that it has been given to people of all ages. The values for COM in Iceland are lower in the second half of the observation period, as the use of AZ and JANSS was quite high, compared to the other countries. AZ was given in Denmark in the first weeks, but they stopped \cite{pottegaard2021arterial}. Furthermore, we see that AZ was not given to the younger age groups in Finland, Iceland nor Sweden. JANSS was given only to a small part of the population in Denmark, and to a larger extent in Iceland, especially to the younger age groups.

\begin{figure}
    \centering
    \includegraphics[scale=0.20]{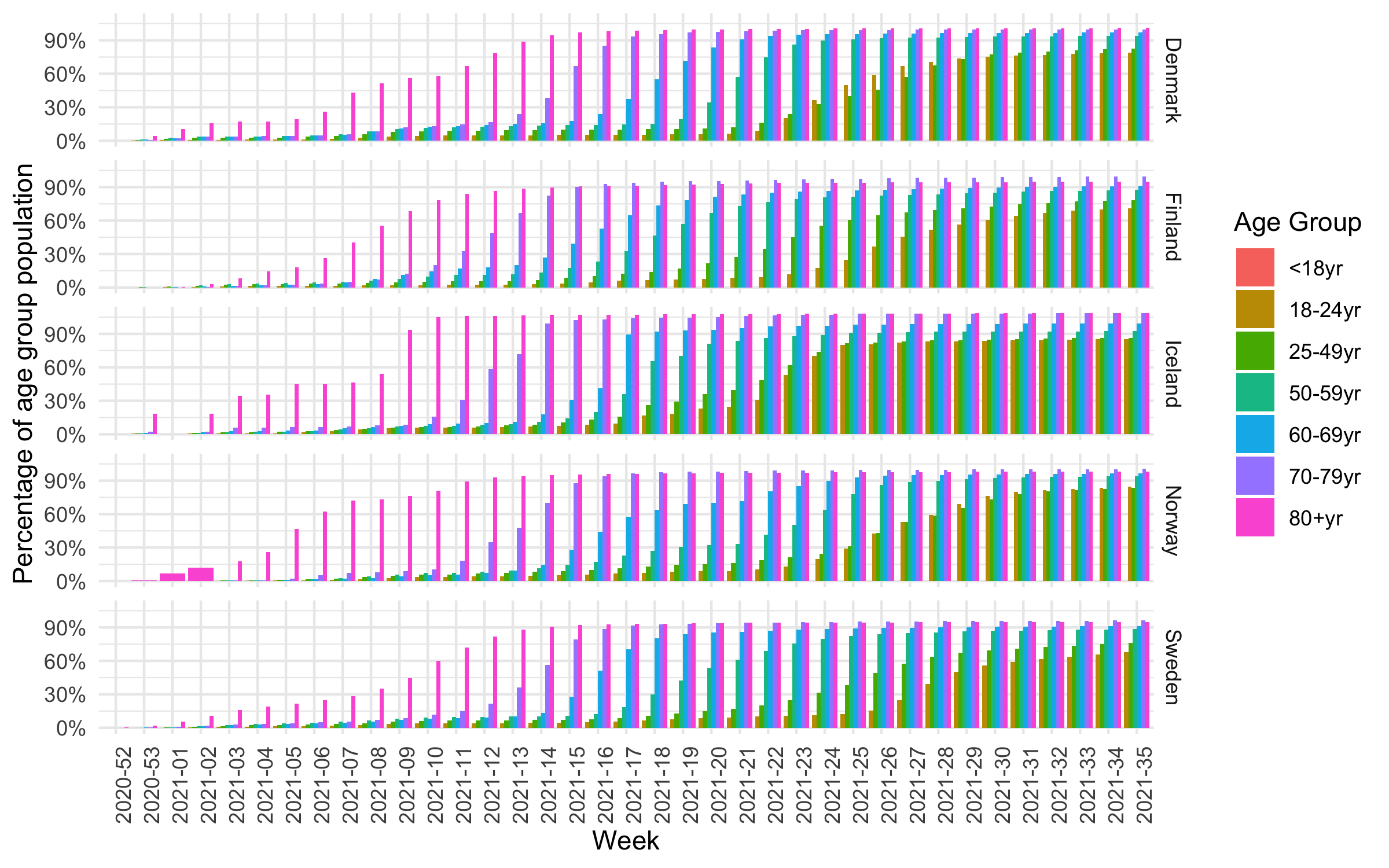}
    \caption{Fraction of population of each age group that has received the first vaccine dose.}
    \label{fig:PercentFirst}
\end{figure}

\begin{figure}
    \centering
    \includegraphics[scale=0.20]{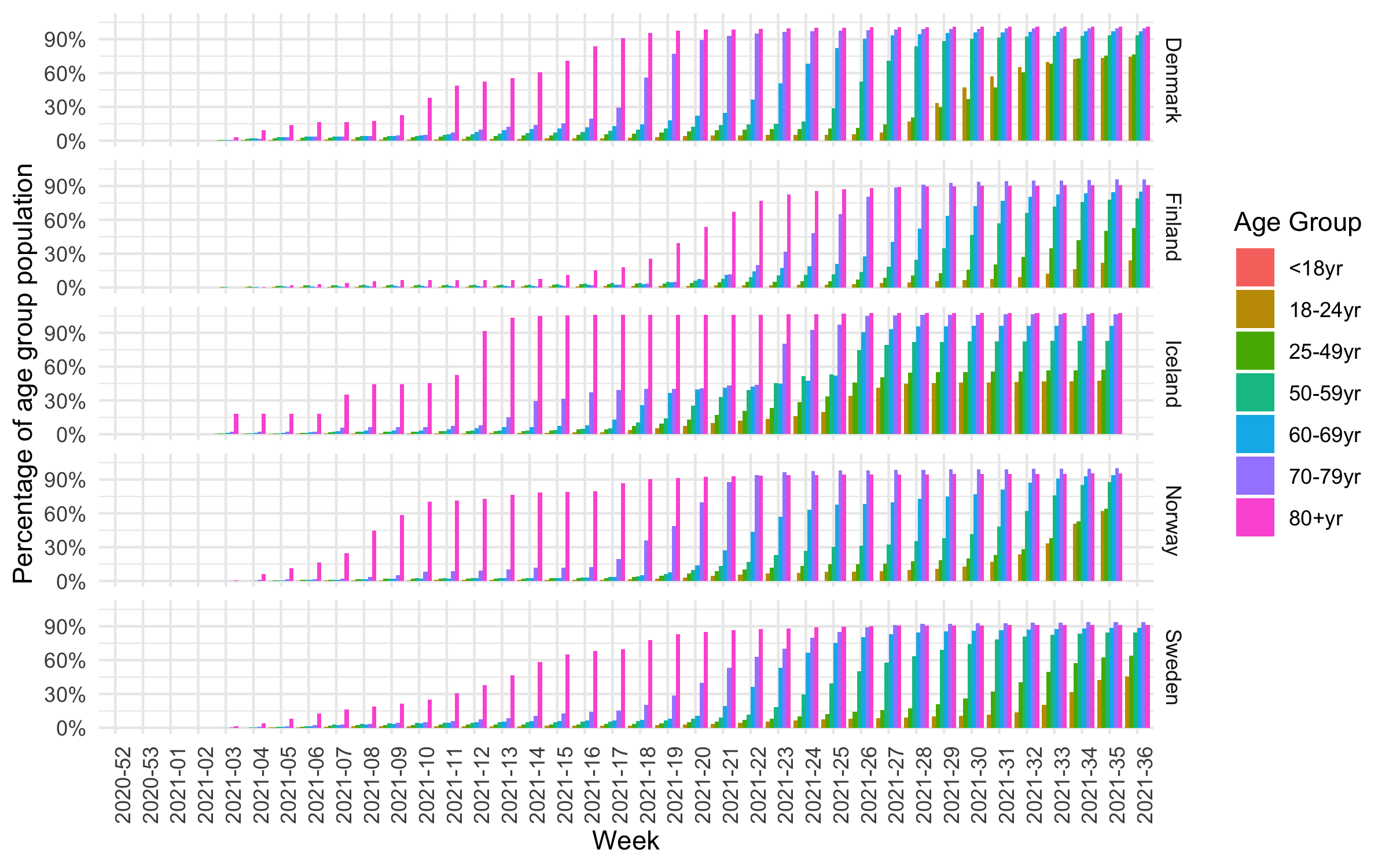}
    \caption{Fraction of population of each age group that has received the second vaccine dose or is fully vaccinated.}
    \label{fig:PercentSecond}
\end{figure}

The last two figures here show the various vaccination strategies in terms of the roll out for the age groups. The effect of the strategies can be seen in Figures \ref{fig:PercentFirst} and \ref{fig:PercentSecond} which show the cumulative percentage uptake for each age group of the first and the second dose respectively.

Although the countries started vaccinating their population around the same time, we can see different trends. All countries focused on the oldest generations in the first few weeks of vaccinations, with some doses given to younger people, which can most likely be explained by healthcare and front-line workers. This is however different in Norway, where only the oldest generation was included in the first three weeks and no significant portion of healthcare workers is detected. Furthermore, Norway also differs as the younger age groups received the vaccination with more delay compared to the other counties. By week 20, Denmark had reached 75\% vaccination uptake for the 60+ age group, at week 19 in Finland, week 17 in Iceland, week 22 in Norway and week 18 in Sweden. 
By the end of the observation period, this goal had not been reached for the $18-24$ age group in Finland and Sweden and for the $25-49$ age group in Sweden.
It was reached in week 33 in Denmark, week 25 in Iceland and week 31 in Norway.

When analyzing the data for the second dose, there is an even greater difference to be observed.
Finland lags behind the other countries when it comes to starting the administration of the second dose.
For the older generations (60+), Denmark, Iceland and Sweden had reach 75\% vaccination uptake for that population in weeks 25 and week 26, whereas Finland and Norway were weeks behind, reaching this milestone in week 31.
By the end of the observation period, Finland, Norway and Sweden are clearly lagging behind in delivering the second dose, especially for people under 50 years old.

\begin{table}[]
   \caption{The  week number in 2021 where at least 10\% of the population within the respective age groups had received the first vaccine dose (1st) and were fully vaccinated (full)  }
    \label{tab:tenPercent}
    \centering
    \begin{tabular}{l|cccccccccccc}
    \toprule
Age Group &\multicolumn{2}{c}{18-24}&\multicolumn{2}{c}{25-49}&\multicolumn{2}{c}{50-59}&\multicolumn{2}{c}{60-69}&\multicolumn{2}{c}{70-79}&\multicolumn{2}{c}{80+}\\
 Uptake      &1st&full&1st&full&1st&full&1st&full&1st&full&1st&full\\
Country&     \\  \midrule
Denmark& W23 &W28 & W16&W22&W9&W19&W9&W14&W9&W12&W1&W5 \\
Finland& W23& W33&W16&W28&W11&W23&W9&W21&W9&W21&W4&W15\\
Iceland& W17&W19&W15&W18&W14&W18&W13&W17&W10&W13&W0&W3\\
Norway&W23&W30&W18&W25&W14&W22&W14&W20&W10&W14&W1&W5 \\
Sweden&W22&W30&W18&W24&W13&W22&W13&W20&W10&W14&W2&W6 \\
EU/EEA    &W20&W25&W16&W22&W12&W19&W12&W18&W10&W14&W4&W8 \\
\bottomrule
    \end{tabular}
\end{table}

Now we can compare the new infection rates to vaccination strategies, by looking at Table \ref{tab:tenPercent}. This table shows the week number in which each country reached a 10\% vaccination level in each age group for the first and second dose respectively. As such, the table illustrates when vaccinations were well underway in each age group and can be seen as a representation of the vaccination strategy.
For example, Iceland was the first country to reach this level in the younger categories for both first dose and for fully vaccinated. To reach this goal, Iceland used all vaccine types, whereas the other countries refrained from using JANSS and AZ (see Table \ref{tab:overview}). Furthermore, Finland reached the 10\% level for the first dose earlier than the other countries, but lagged behind for the second dose, indicating that Finland prioritized the administration of the first dose to a large part of the population instead of fully vaccinating a smaller portion of the population.

The impact of the vaccination strategies on new infections is clear.
Finland has a larger wave at the end of the observation period, a wave which is especially visible among the younger generations. Finland also has a more severe wave, with less reduction in the number of hospitalizations and deaths, compared to the last wave and the other countries.

However, in all countries, the new wave which starts in the summer months of 2021 is dominated by the younger age groups who are getting infected to a larger extent compared to the older age groups.

\subsection*{Rolling Correlation}\label{sec:Rolling}
\begin{figure}[!ht]
\centering
\subfloat[]{\includegraphics[width=0.18\linewidth]{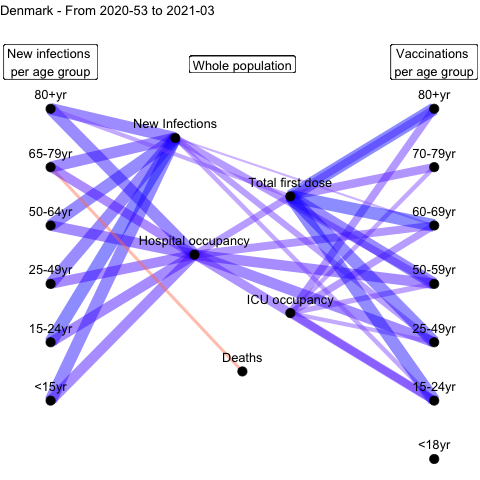}\label{denmark1}}
\hfil
    \subfloat[]{\includegraphics[width=0.18\linewidth]{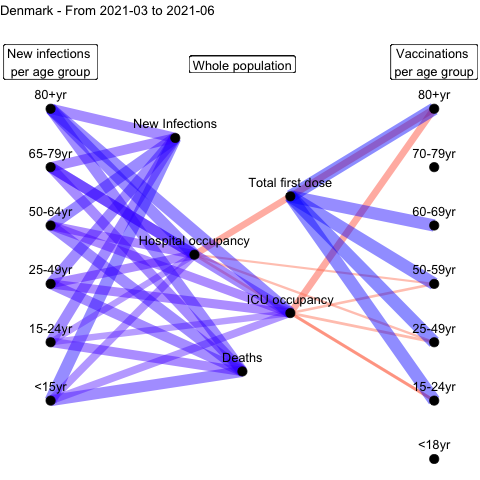}\label{denmark2}}
\hfil
    \subfloat[]{\includegraphics[width=0.18\linewidth]{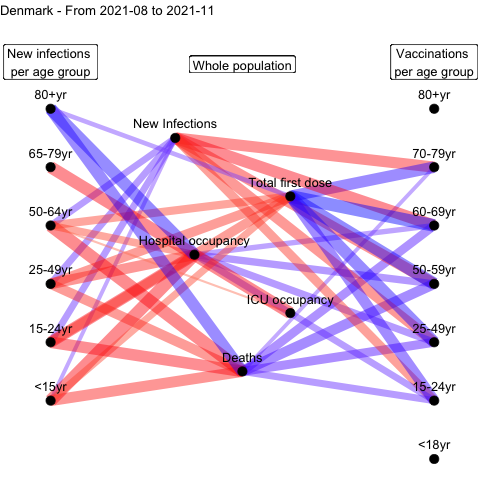}}
\hfil
    \subfloat[]{\includegraphics[width=0.18\linewidth]{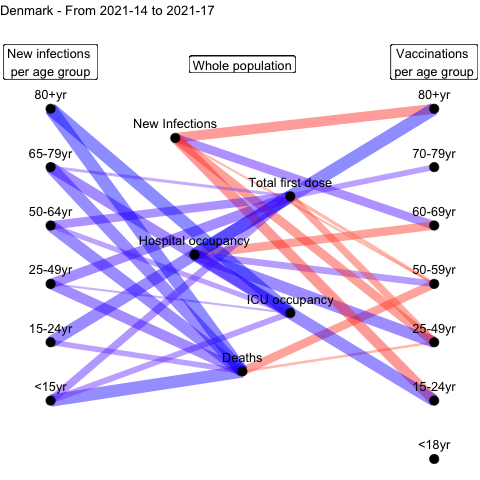}}
    \hfil
    \subfloat[]{\includegraphics[width=0.18\linewidth]{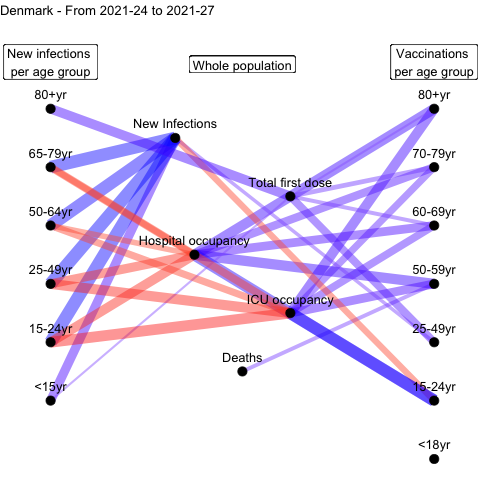}}

 \subfloat[]{\includegraphics[width=0.18\linewidth]{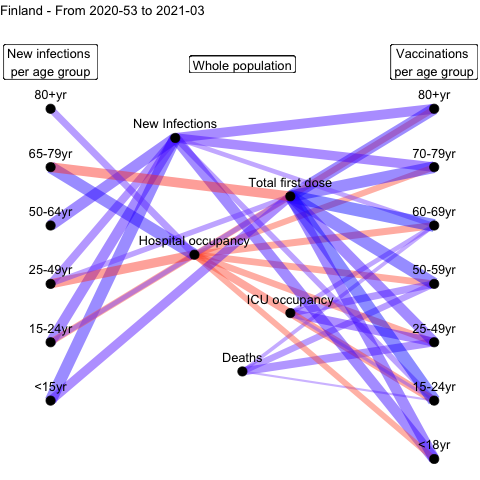}\label{finland1}}
\hfil
    \subfloat[]{\includegraphics[width=0.18\linewidth]{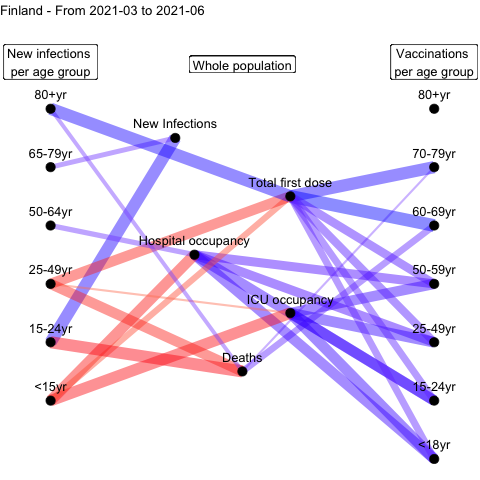}}
\hfil
    \subfloat[]{\includegraphics[width=0.18\linewidth]{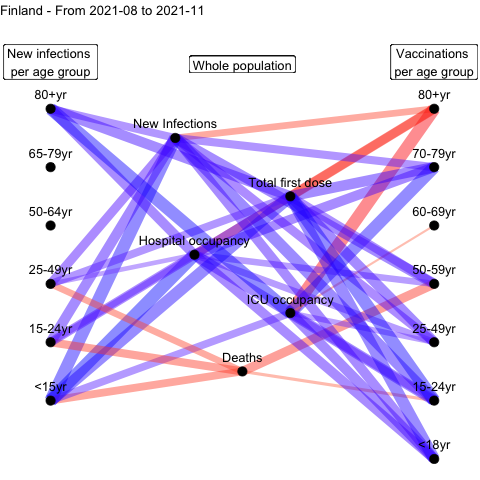}}
\hfil
    \subfloat[]{\includegraphics[width=0.18\linewidth]{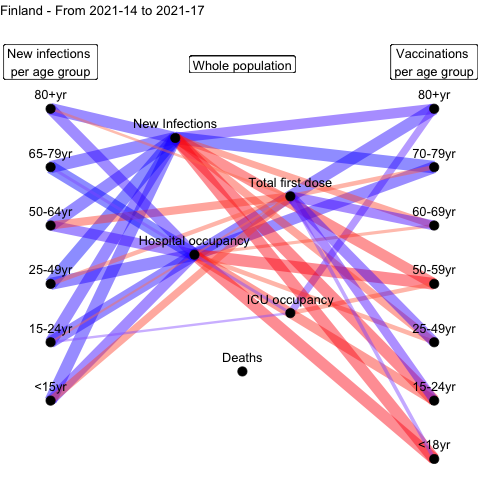}}
\hfil
    \subfloat[]{\includegraphics[width=0.18\linewidth]{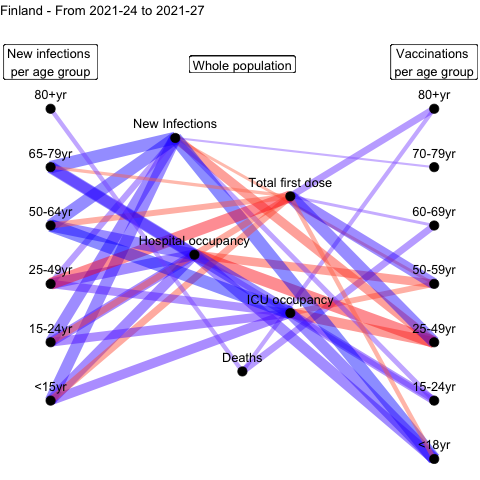}\label{finland5}}

 \subfloat[]{\includegraphics[width=0.18\linewidth]{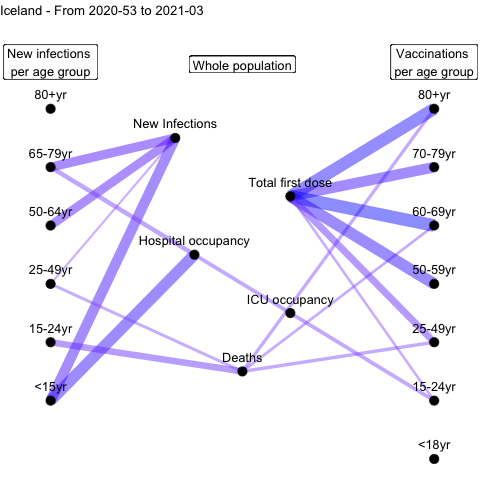}}
\hfil
    \subfloat[]{\includegraphics[width=0.18\linewidth]{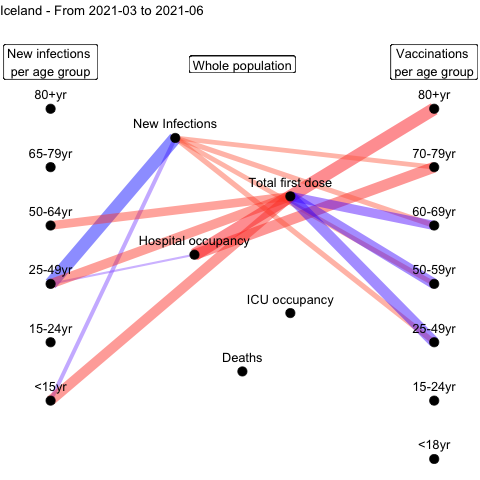}}
\hfil
    \subfloat[]{\includegraphics[width=0.18\linewidth]{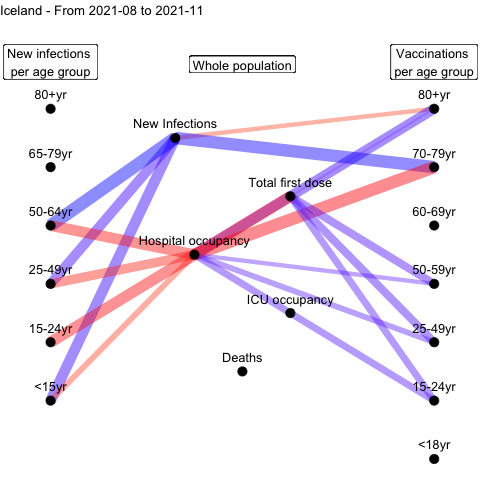}}
\hfil
    \subfloat[]{\includegraphics[width=0.18\linewidth]{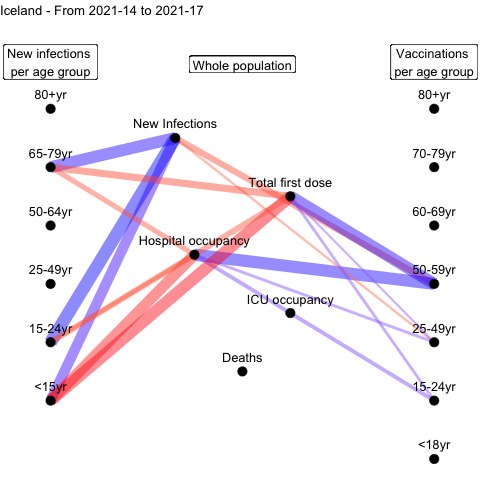}}
    \hfil
    \subfloat[]{\includegraphics[width=0.18\linewidth]{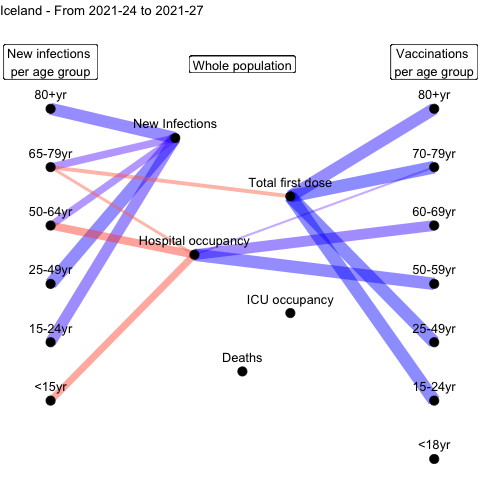}}

 \subfloat[]{\includegraphics[width=0.18\linewidth]{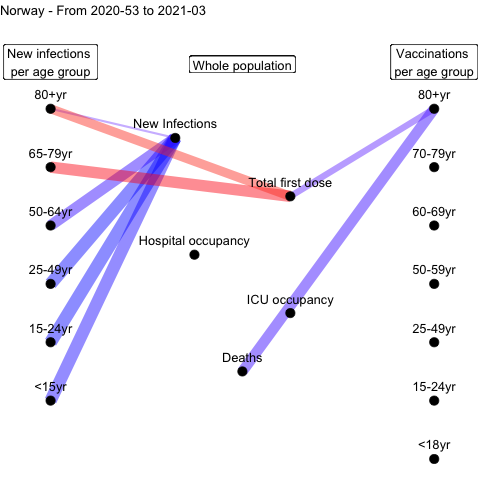}}
\hfil
    \subfloat[]{\includegraphics[width=0.18\linewidth]{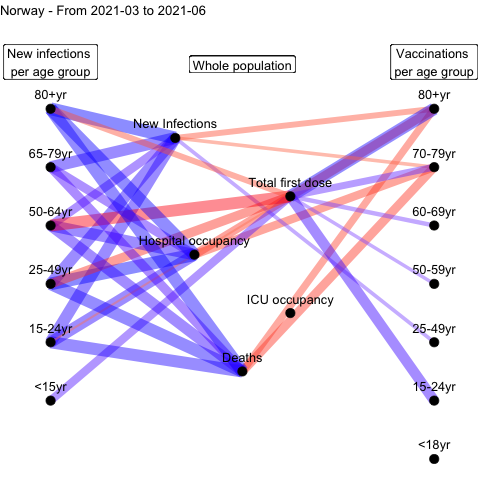}}
\hfil
    \subfloat[]{\includegraphics[width=0.18\linewidth]{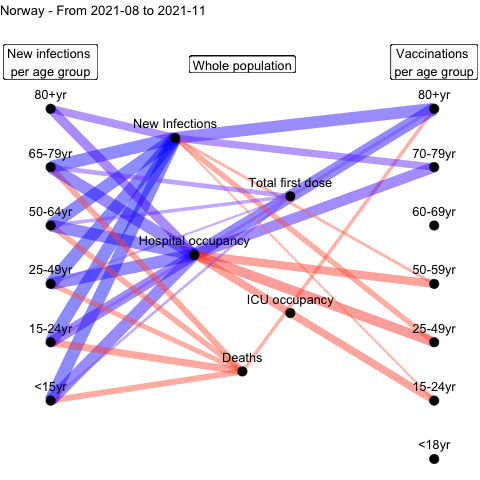}}
\hfil
    \subfloat[]{\includegraphics[width=0.18\linewidth]{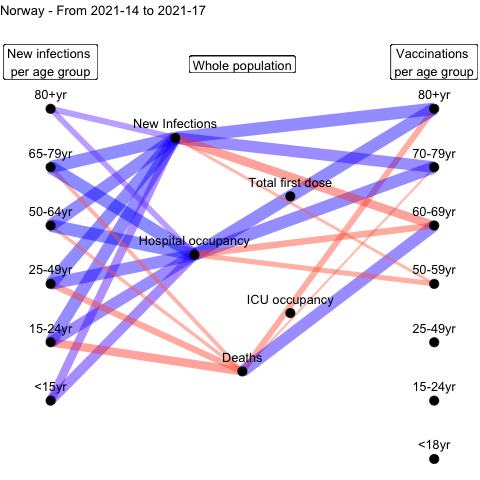}}
    \hfil
    \subfloat[]{\includegraphics[width=0.18\linewidth]{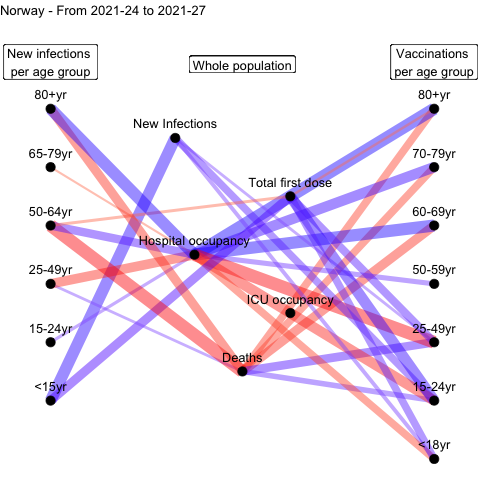}}
    
    \subfloat[]{\includegraphics[width=0.18\linewidth]{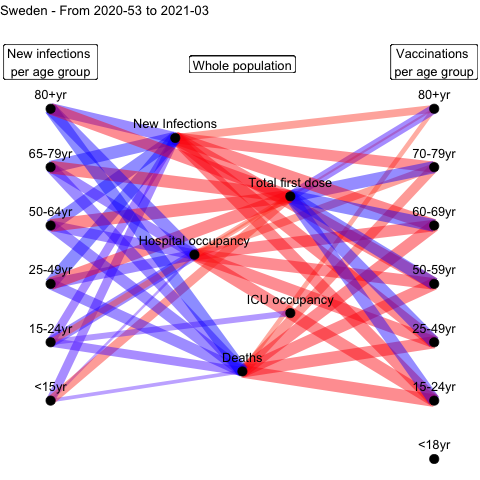}\label{sweden1}}
\hfil
    \subfloat[]{\includegraphics[width=0.18\linewidth]{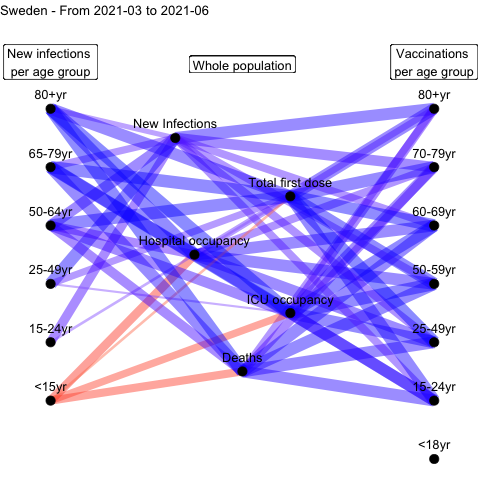}\label{sweden2}}
\hfil
    \subfloat[]{\includegraphics[width=0.18\linewidth]{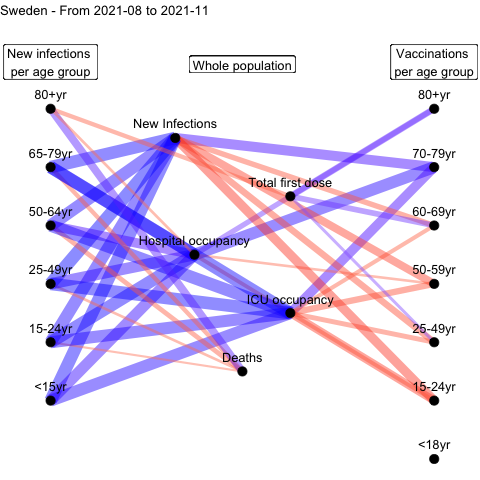}}
\hfil
    \subfloat[]{\includegraphics[width=0.18\linewidth]{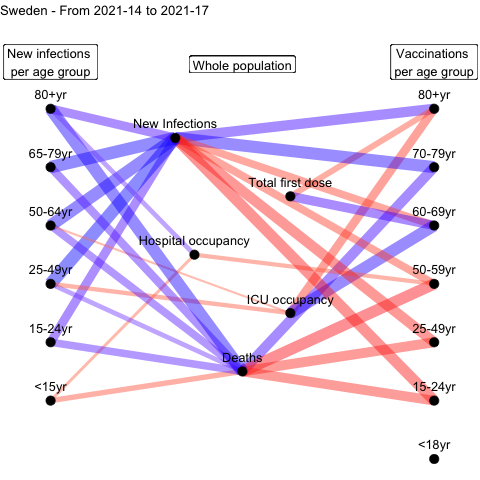}}
\hfil
    \subfloat[]{\includegraphics[width=0.18\linewidth]{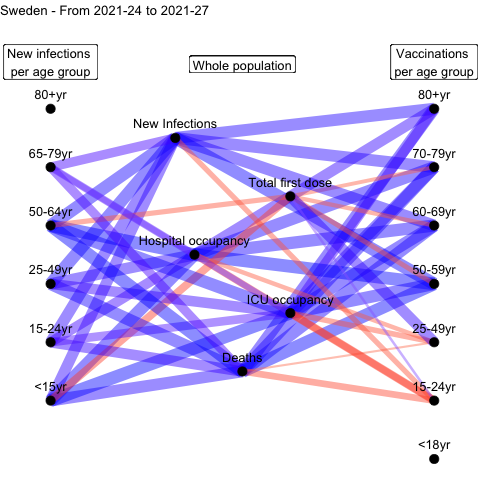}}
\caption{Rolling correlations between changes in population indicators (middle) and infections in age groups (left) and vaccination of age groups (right). Rows from top to bottom: Denmark, Finland, Iceland, Norway, Sweden. Column left to right: From 2020-53 to 2021-03,From 2021-03 to 2021-06,From 2021-08 to 2021-11,From 2021-14 to 2021-17,From 2021-24 to 2021-27. Blue indicates a positive correlation and red a negative correlation, and a thicker edge indicates a stronger correlation. Correlations with an absolute value below 0.6 are omitted.}
    \label{fig:Rolling}
    \end{figure}

Now we look at the rolling correlations to inspect trends in the relationships between variables.
Figure \ref{fig:Rolling} shows the rolling correlations for five periods in time, in the five Nordic countries. Each figure has three parts: population indicators (middle), infections in age groups (left) and vaccination of age groups (right). The five periods are: i) Weeks 2020-53 to 2021-03, ii) Weeks 2021-03 to 2021-06, iii) Weeks 2021-08 to 2021-11, iv) Weeks 2021-14 to 2021-17, v) Weeks 2021-24 to 2021-27. 
The left side of each of these figures shows the progression of the pandemic, in terms of which age groups are getting infected and hospitalized, where as the right hand side shows the vaccination strategy. 

We see interesting trends for all countries. Denmark and Sweden show similar patterns at the start of the observation period, see Figures \ref{denmark1} and \ref{denmark2} for Denmark and \ref{sweden1} and \ref{sweden2} for Sweden. As is visible in Figure \ref{fig:Per100kAgeGroup}, the wave is going down and so are the number of infections and hospitalizations, as indicated by the positive correlations shown here. However, in Denmark about the same number of people were vaccinated in all age groups about the same number of people per week in the beginning, while in Sweden the number grew every week during the first four weeks of vaccinations. This is why there is a negative correlation between the vaccinations per age group and the indicators of the whole population in Figure \ref{sweden1}. 

In Sweden, infections of all age groups are positively correlated with new infections in the beginning. Looking towards the seriousness of the illness, we see that all age groups are positively correlated with hospital occupancy and deaths, whereas only people aged 18-24 years are correlated with ICU occupancy. In contrast to figure \ref{fig:Per100kAgeGroup}, infections, deaths and hospital occupancy show a negative trend in the first four weeks, while the ICU occupancy is more or less stable. Moving on to weeks 3 to 6, the highest five age groups have a positive correlation with new infections; indeed the number of infections are increasing for all of these age groups.

In Finland, at the end of the observation period (Figure \ref{finland5}, all age groups except the oldest have a high positive correlation with new infection, whereas only the younger groups ($<15$, $15-24$,$24-49$ and $50-64$ are correlated with hospitalizations and ICU occupancy, indicating that these are the age groups with more severe illness.  

Norway looks different from the other observed countries when it comes to the vaccination correlations. For each period, there are fewer positive correlations between vaccinated age groups and total vaccines administered compared to what we observe for the other countries. This might indicate that the vaccination strategy was more divided and that the focus was to finish vaccinating one age group before starting the next one.

The number of new infections in each age group is positively correlated with total number of new infections in variable ways. In these 25 figures we get a positive correlation between new infections:  
12 times for the oldest age groups ($80+$), 17 times for the second highest ($65-79$) and 19 times or more for the remaining age groups. This means that the oldest generations that were vaccinated first seem to be better protected when it comes to new infections.

\subsection*{eRG results} 
\begin{figure}[h!]
    \centering
    \includegraphics[scale=0.5]{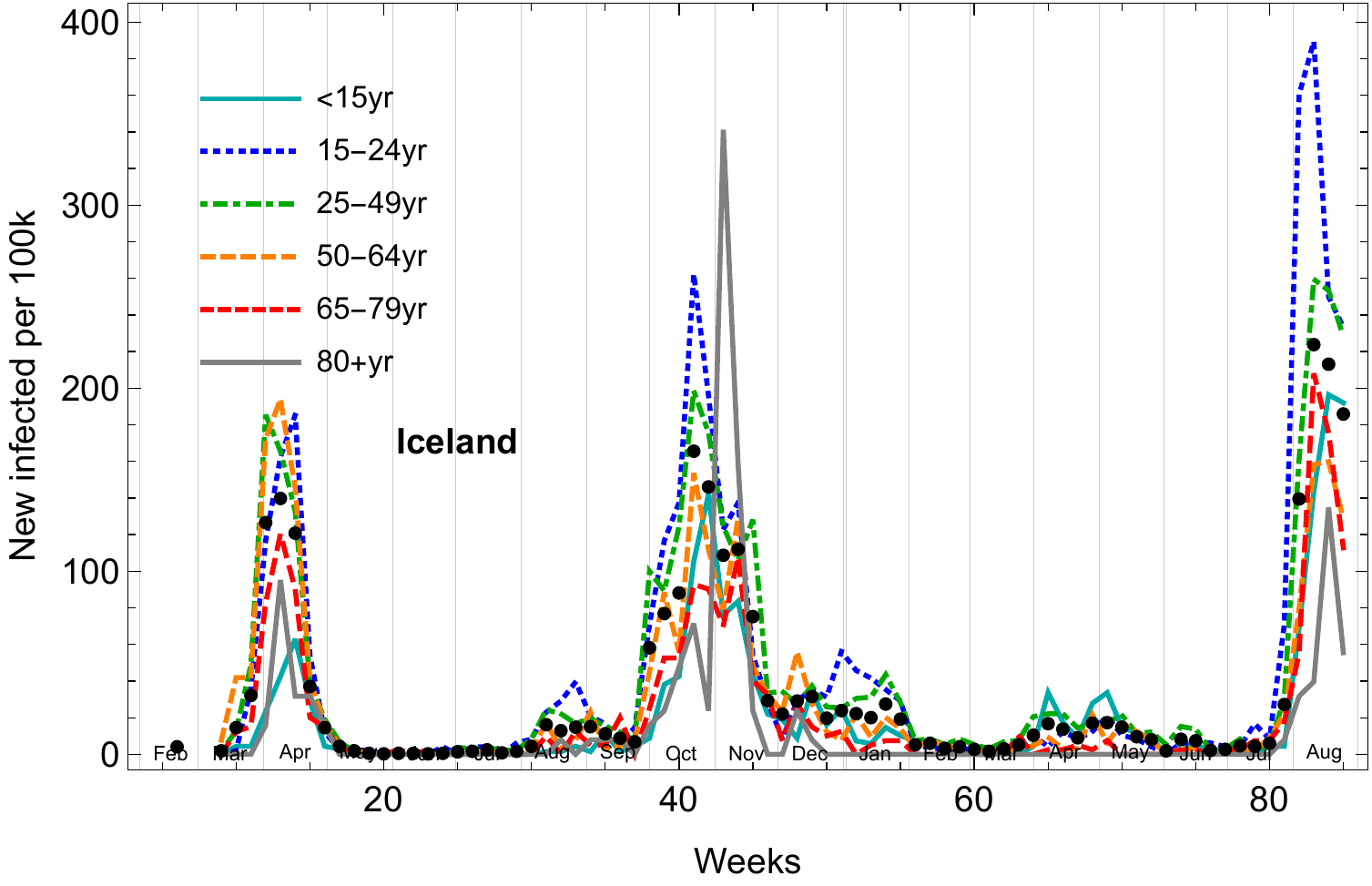}
    \includegraphics[scale=0.5]{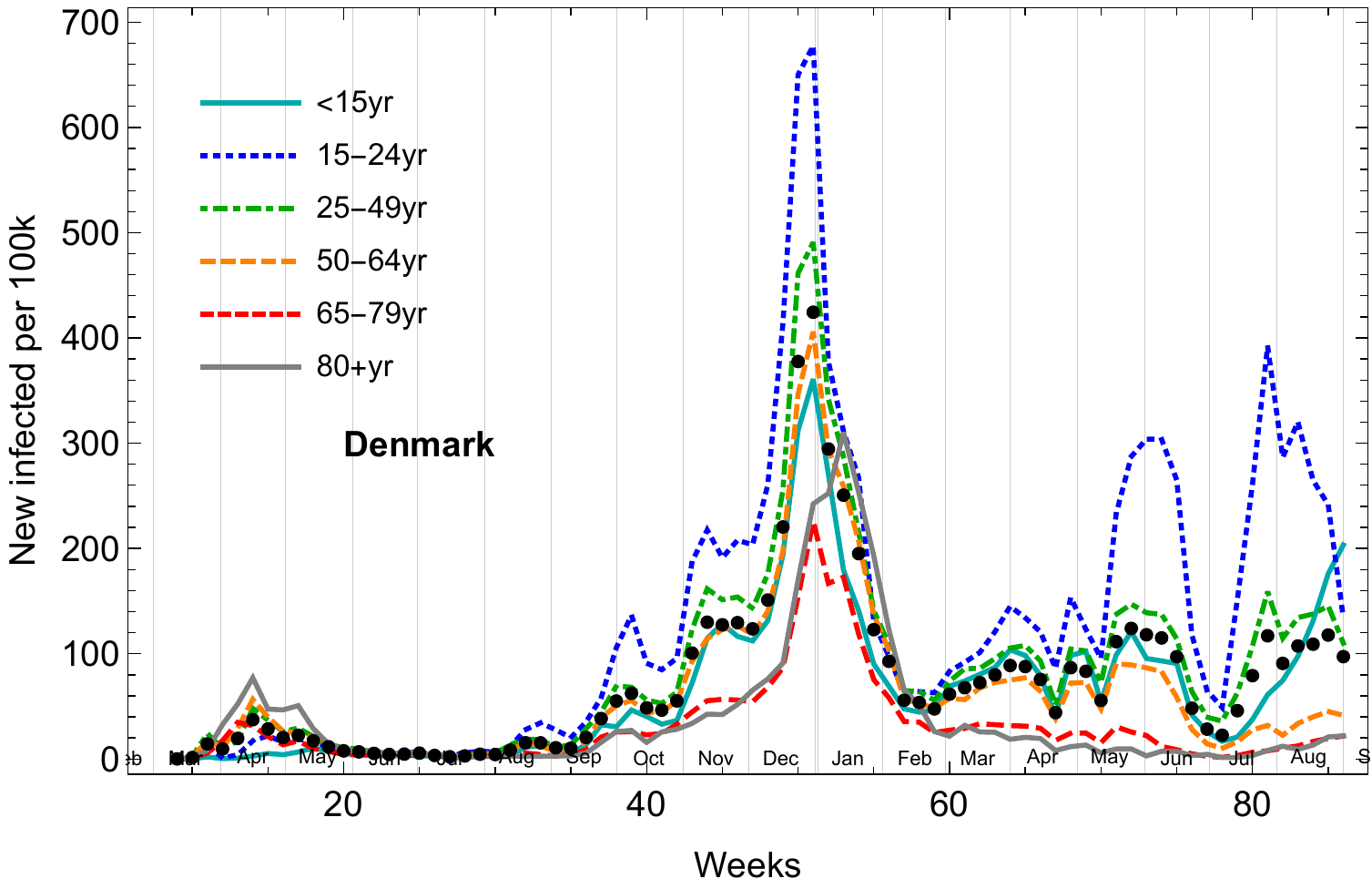}
    \includegraphics[scale=0.5]{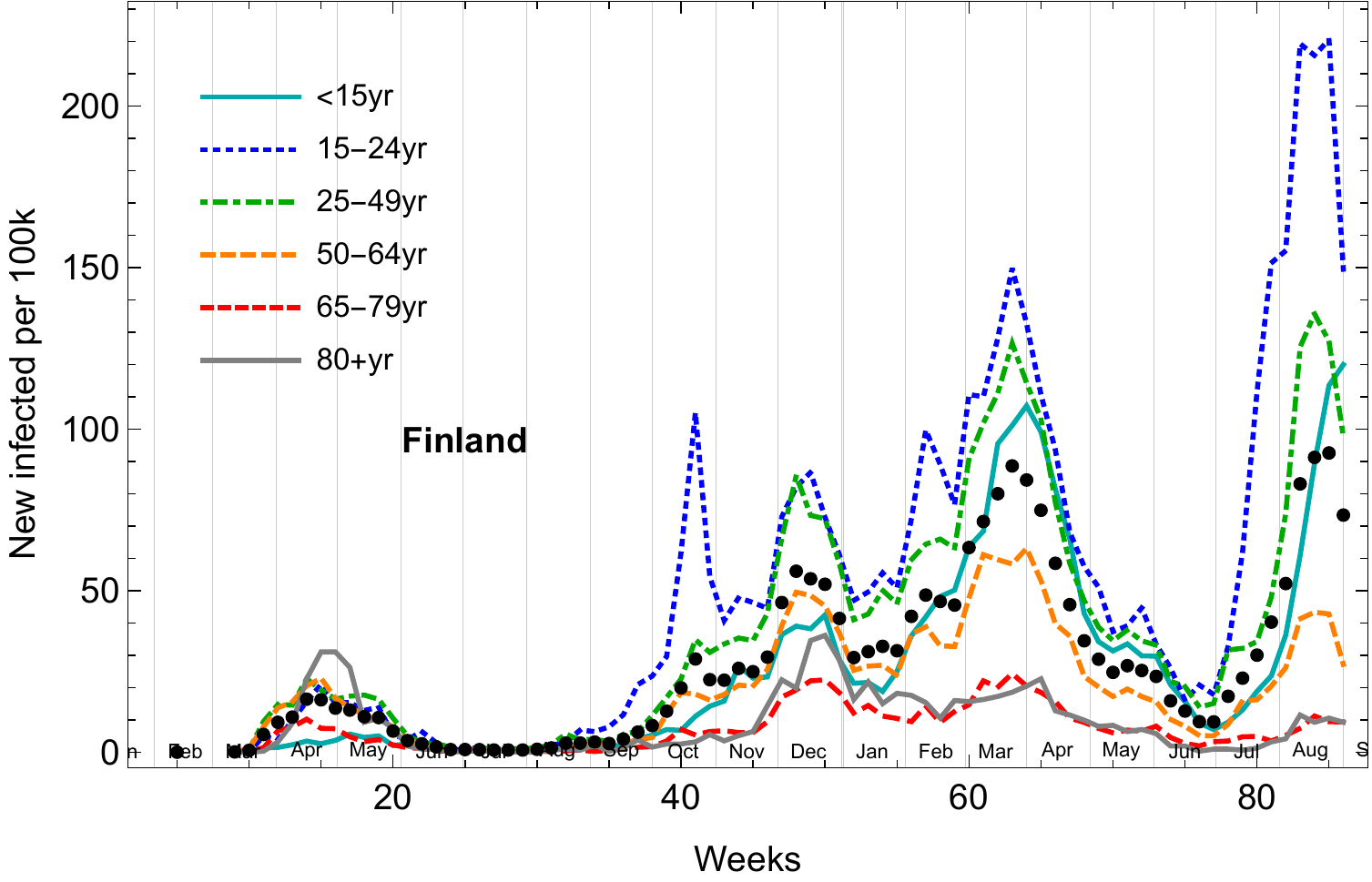}
    \includegraphics[scale=0.5]{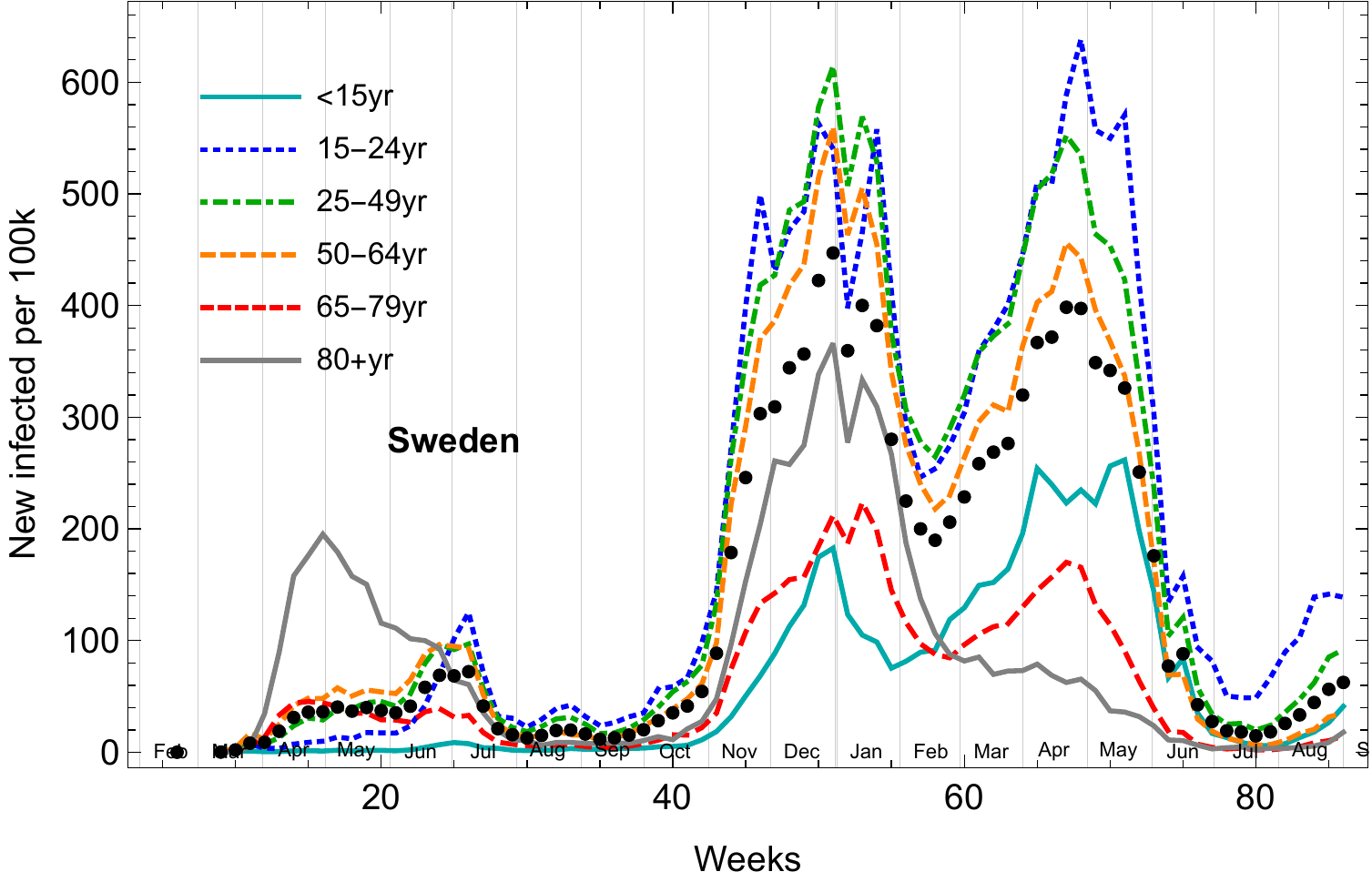}
    \includegraphics[scale=0.5]{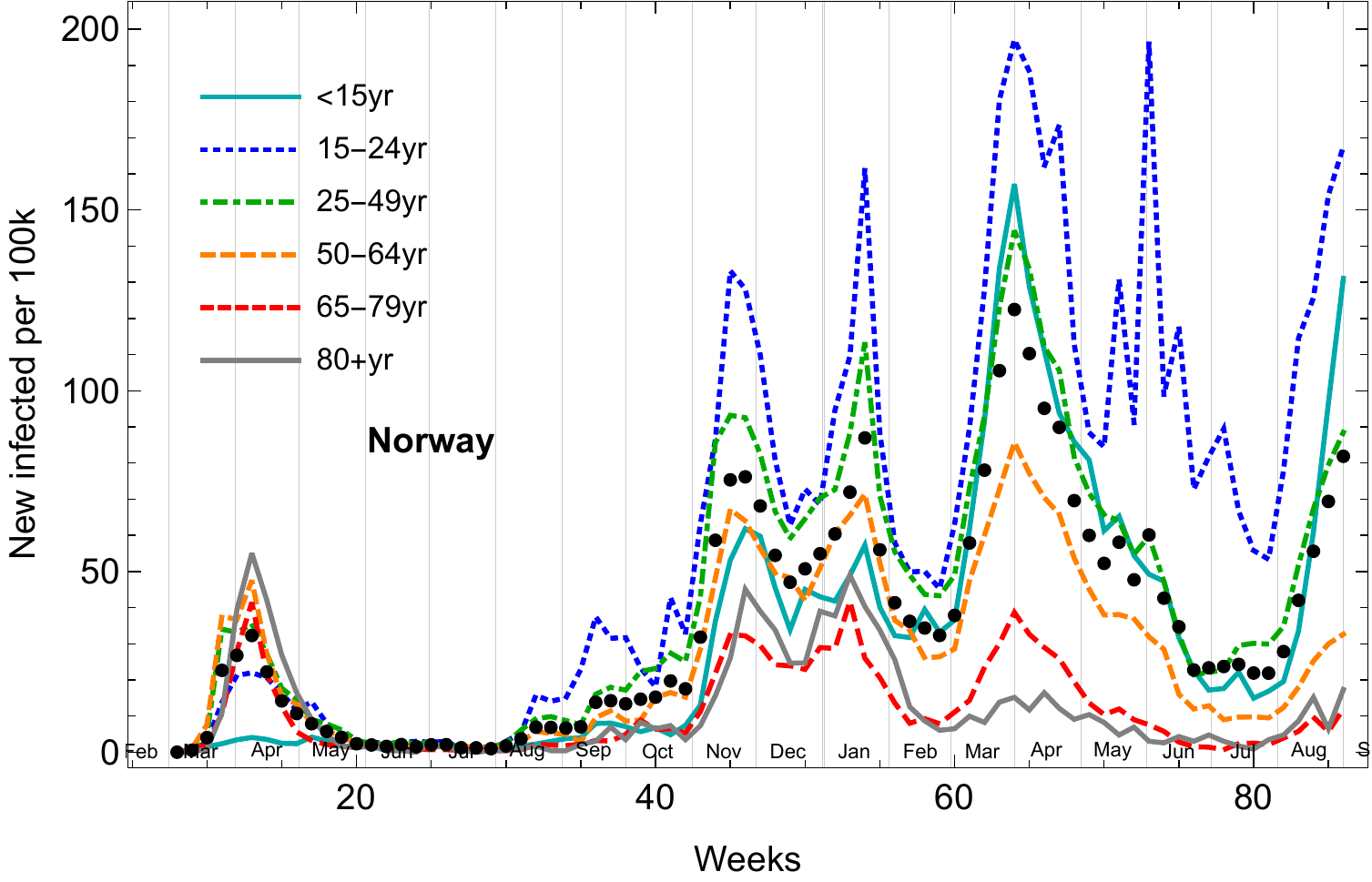}
    \\
    \includegraphics[scale=0.5]{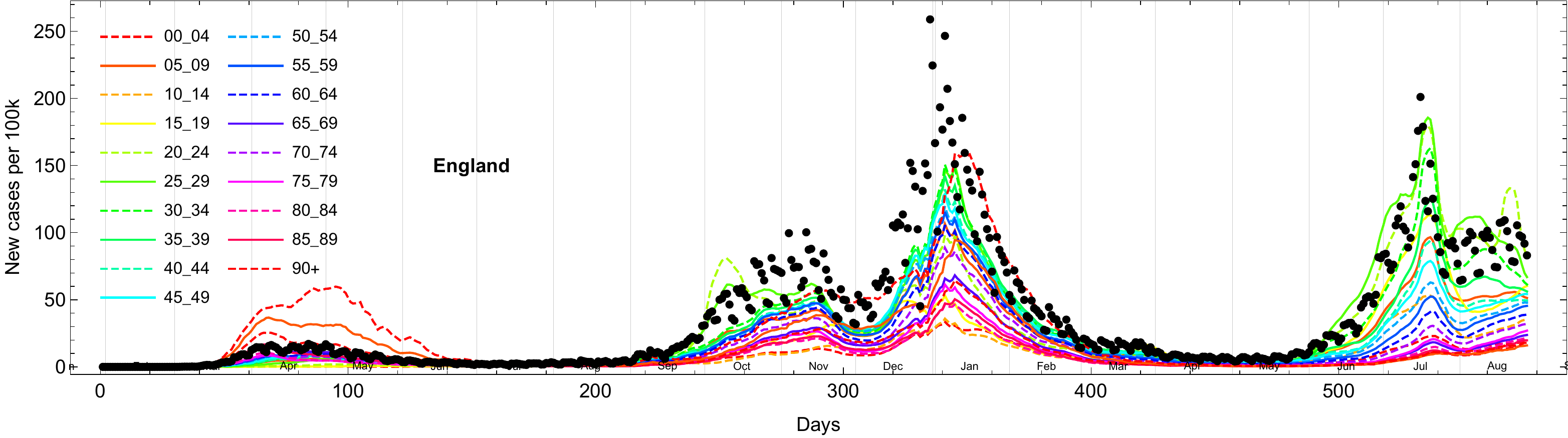}
    \caption{Number of cases per 100k inhabitants from each age class within Iceland, Denmark, Finland, Sweden, Norway and England. Black dots represents the number of new cases per 100k inhabitants for all age groups.}
    \label{fig:New_cases}
\end{figure}

\begin{figure}
    \centering
    \includegraphics[scale=0.55]{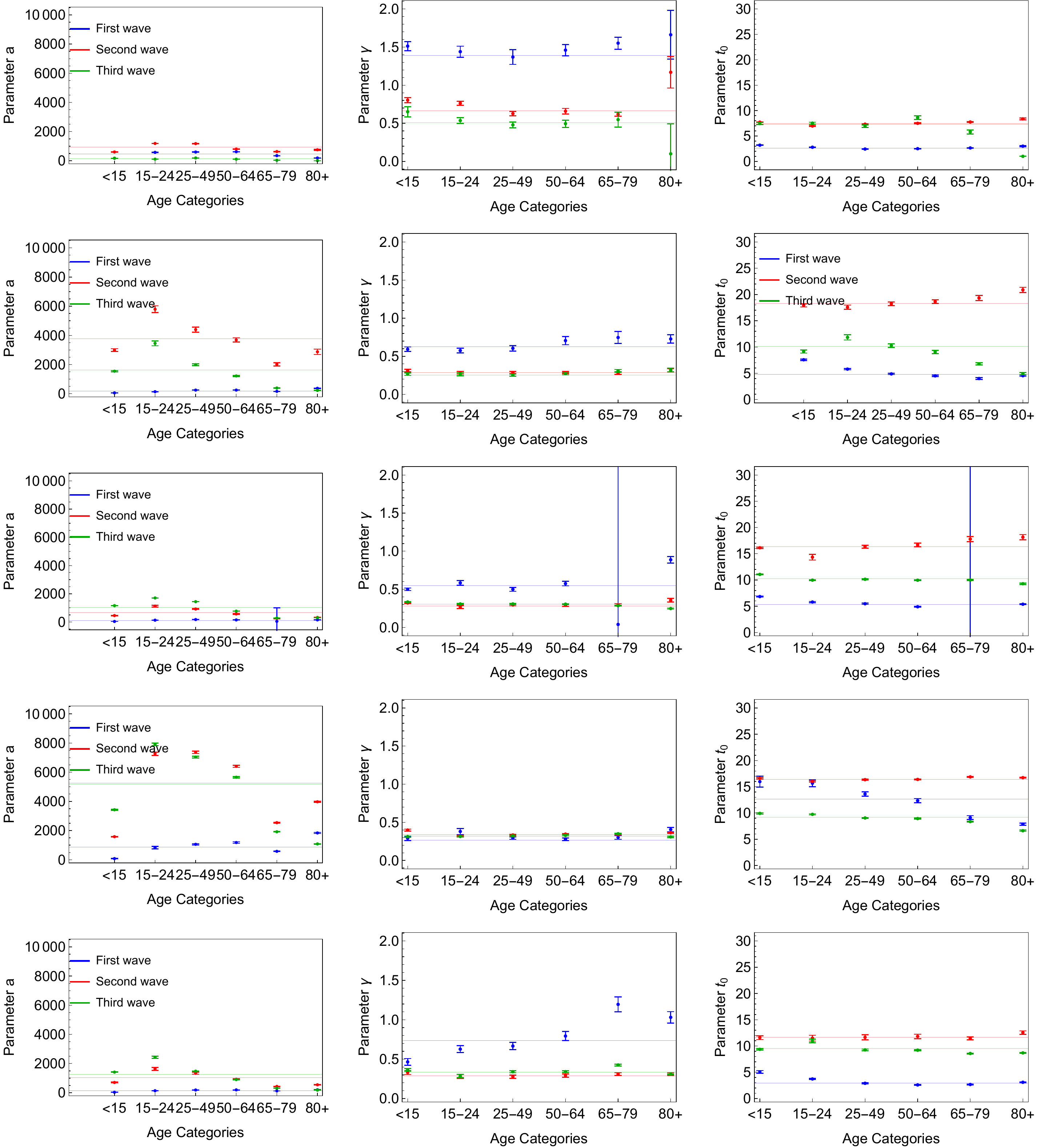}
    \caption{Fit parameters for Iceland, Denmark, Finland, Sweden and Norway}
    \label{fig:fitparameters}
\end{figure}

Now we would like to study the evolution of the epidemic waves for each age group in the Nordic countries through the scope of the epidemic Renormalisation Group (eRG) formalism. First, when examining the data for the number of new cases per 100k inhabitants from each age group (see Figure \ref{fig:New_cases}), the 80+ category behaves differently, across the different waves. We can see that the 80+ age group has fewer cases for the later waves and that trend is increasingly visible after the beginning of the vaccination campaigns early 2021. Furthermore, we observe that the 15-24 age group represents the major part of the new cases reported while being one of the least vaccinated group.
One way to quantify these observations is to study the data through the eRG formalism \cite{10.3389/fphy.2020.00144} and the logistic function:
\begin{equation}
    I(t) = \frac{ae^{\gamma (t-t_0)}}{1 + e^{\gamma (t-t_0)}}
\end{equation}
Each age group, wave or country is then comparable with another for both the total number of cases (per 100k inhabitant) per wave "$a$" and the infection rate $\gamma$ as can be seen in Figure \ref{fig:fitparameters}. In this part of the analysis, we also take into account a time shifting parameter $t_0$ representing the relative timing of the peak among the age groups for each wave.
First, we can see that the parameter $a$ is always smaller for the first wave compared to the next ones. This observation can be explained by the testing capacity difference between the beginning of the COVID-19 outbreak across the world, compared to later times. A lower number of tests taken logically brings a lower number of reported cases and less relevance of the normalisation of the number of cases. This may also explain the over-representation of the 80+ age group during the first wave, as retirement homes were tested to a larger extent in some of the observed countries. However, the comparison between the parameter $a$ within each wave shows that the parameter is smaller for higher age groups for the third wave. This may be a direct effect and reflection of the vaccination strategy to vaccinate mostly the highest age groups first.
\par
\begin{figure}
    \centering
    \includegraphics[scale=0.55]{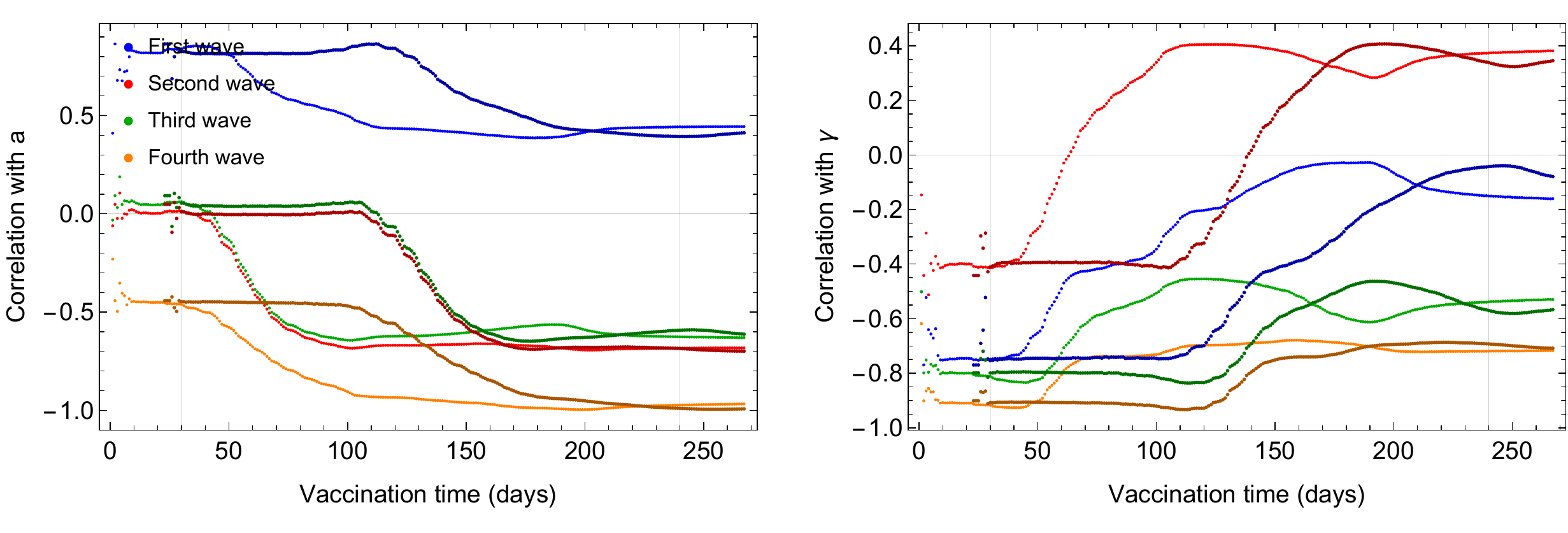}
    \caption{Correlation between vaccination and eRG parameters for the last wave in England. The vertical grey lines indicate the peak timing of the two last waves in England.}
    \label{fig:Correlation-England}
\end{figure}

\begin{figure}
    \centering
    \includegraphics[scale=0.5]{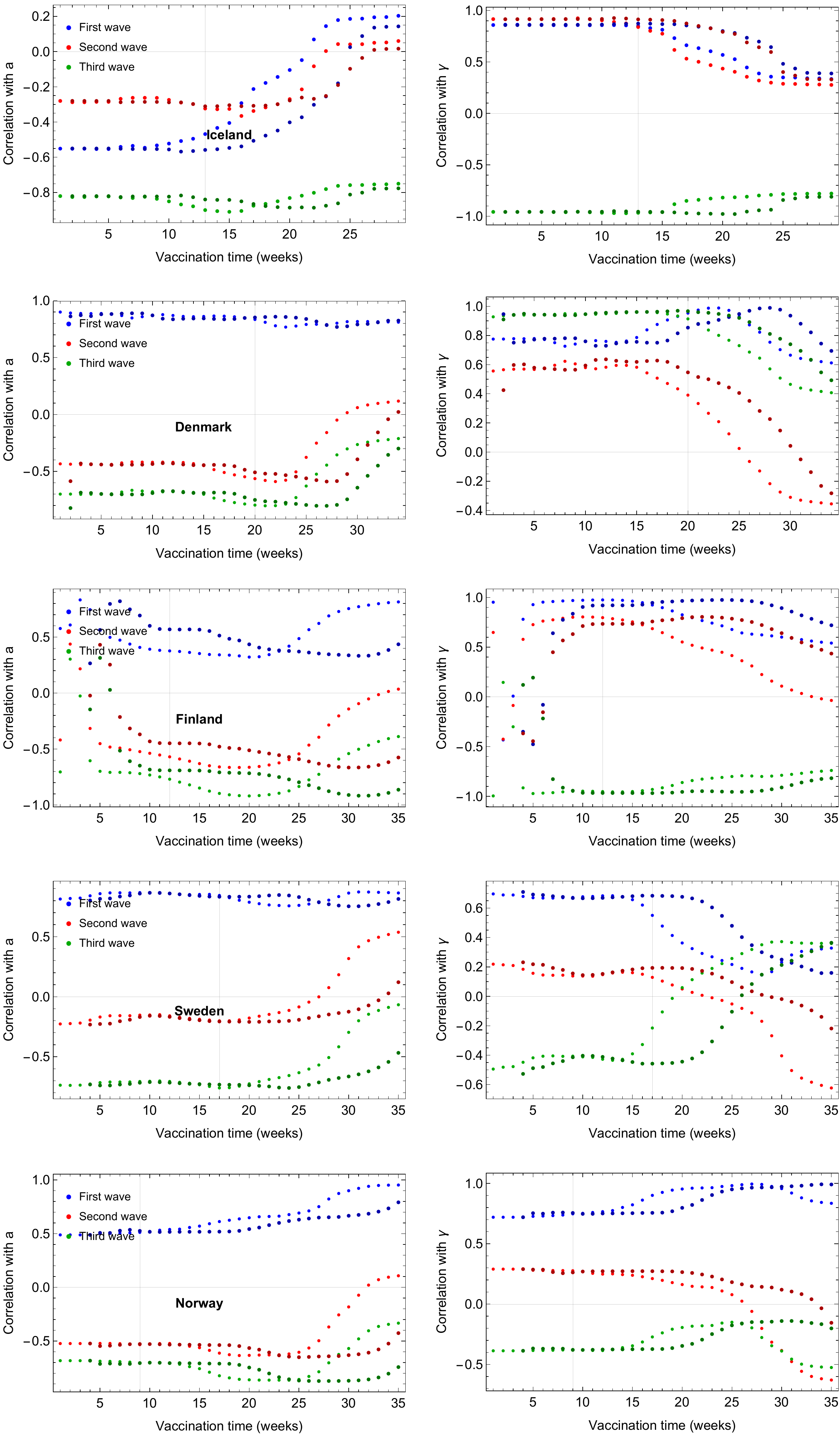}
    \caption{Correlation between the last wave fit parameters and vaccination percentage for Iceland, Denmark, Finland, Sweden and Norway. The vertical grey line indicates the peak timing of the last wave in the country.}
    \label{fig:VaxCorfitparameters}
\end{figure}

 To further validate the action of the vaccination strategies on the evolution of the pandemic, we also examined any correlation between the vaccination percentage and the eRG parameters. The eRG parameters integrate the time evolution of the whole wave while the vaccination rates have been changing over time. To take this into account, therefore, we show in Figures \ref{fig:Correlation-England} and \ref{fig:VaxCorfitparameters} the rolling correlation between the time series of the vaccination percentage per week, since the beginning of the campaign and the fixed eRG parameters for the relevant wave. From our previous works \cite{cot2021impact}, we know that only the vaccination before the peak has a measurable impact on the wave. As a sanity check, we report the correlation with the eRG parameters of previous waves (two in the Nordic countries and three in England) where no meaningful correlation should be observed. If any correlation exists, one could expect that it would be seen only for the last wave, which takes place after the vaccination campaigns begun, and close to the peak of the wave. We expect this correlation to be close to minus one, as more vaccinated individuals bring less infections, so a smaller $a$ and a smaller $\gamma$, while correlation for other waves should be random. The results shown in Figure \ref{fig:VaxCorfitparameters} are, therefore, coherent for parameters $a$ and $\gamma$ and for almost any country. However, a problem seems to appear for $\gamma$ in Denmark, but when looking at the behavior of the third wave in Denmark from Figure \ref{fig:New_cases}, one can see that there was hardly any rise in the wave for older age groups; we therefore conclude that the fit is sub-optimal here.

\begin{figure}
    \centering
    \includegraphics[scale=0.8]{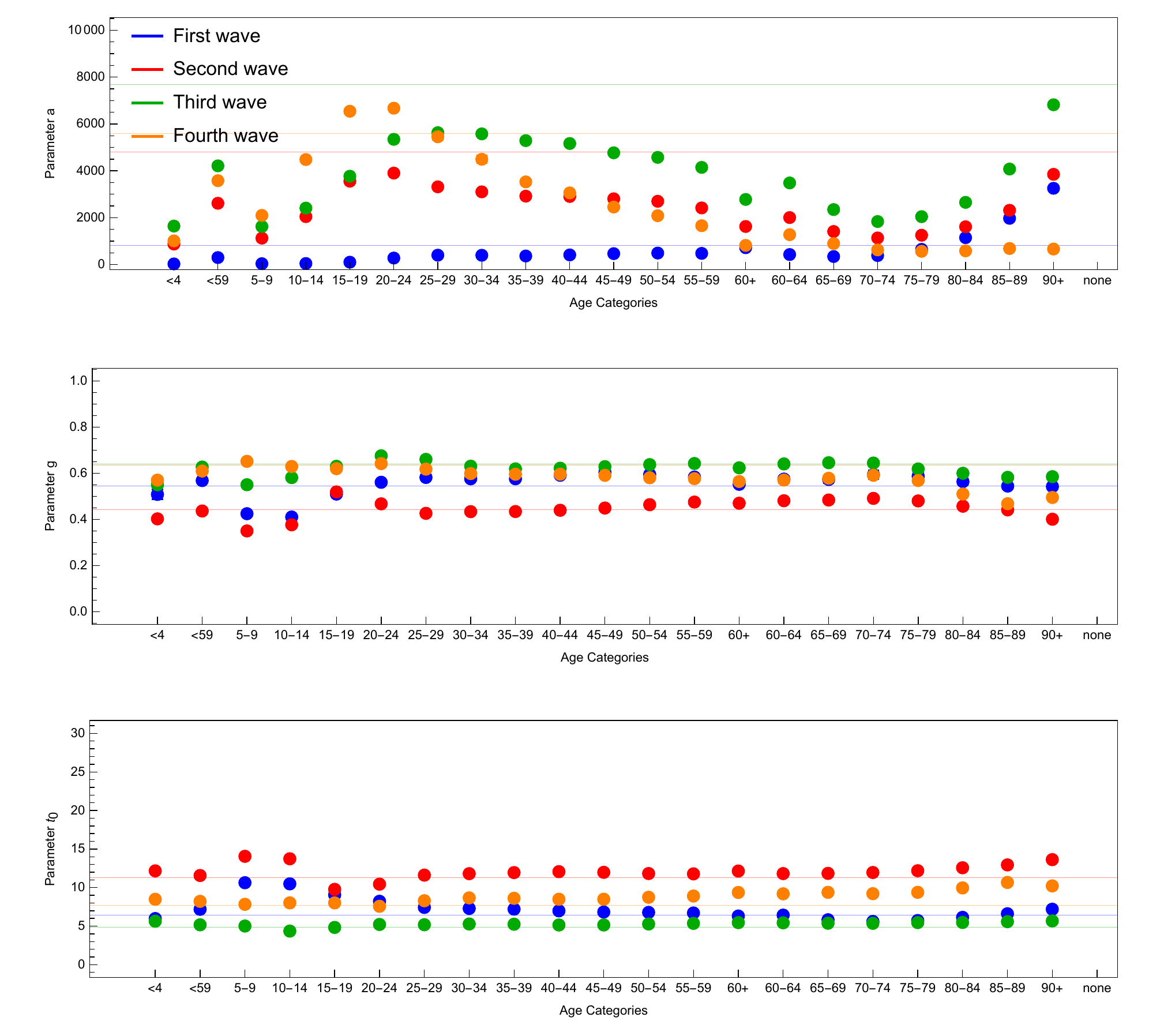}
    \caption{Fit parameters for England.}
    \label{fig:fitparameters-England}
\end{figure}

We would like to compare our previously outlined results for the Nordic countries to England. We chose England because much like the Nordic countries which we have focused on, England had high vaccination uptake early. Observing the fit parameters for England, the vaccination campaign started early and the fourth wave took place 7-8 months after the vaccination started when compared to the start of the vaccination. The population was highly vaccinated in April 2021 (the plateau of vaccinations was reached for the 70+ age group then). In Figure \ref{fig:fitparameters-England}, we can see the fit parameters for the different age groups and waves in England. What can be observed here is that the distribution for $a$ changes significantly for the fourth wave, where older age groups have a smaller parameter, while the age groups for 10-24 years old seem to have an high number of new infections. The shrink for older age groups can be explained by the high number of vaccinations administered, while the rise for younger age groups can be explained by the delta variant that was the primary variant and source for infections during the fourth wave in combination with comparatively lower vaccination uptake for those age groups. 

We would now like to turn the attention towards the correlation between vaccination percentage and eRG parameters among the different age groups (see Figure \ref{fig:Correlation-England}). Here we observe that the fourth wave eRG parameters have a clear anti-correlation with the vaccination percentage, and similarly for $\gamma$. 
\\
The results further corroborate the findings in our previous work \cite{cot2021impact} about the US vaccination campaign impact on the epidemiological data. Indeed, the most vaccinated age groups have a significant reduction in $a$ and $\gamma$.

\section*{Discussion}
In this paper we examine the deployment of vaccines in the Nordic countries in a comparative analysis. Our main results are the quantification of the impact of the vaccination campaigns on age groups through epidemiological data, across countries with high vaccine uptake. Through our findings, we clearly show that the vaccinations have hindered the spread of the virus.

The full extent of the healthcare stress level is difficult to predict until the pandemic has subsided \cite{carda2020role}. There have been attempts on assessing the impact on a micro level but in this paper we analyze the impact on a meso and macro level instead; through data. We analyze the severity and healthcare stress levels through data on new infections, hospitalizations, intensive care unit (ICU) occupancy and deaths. What can be derived from the results is that healthcare stress level has been high, i.e., that most countries have been operating at capacity when it comes to hospital resources at different time periods during the COVID-19 pandemic. 

For that purpose, we consider key indicators of the severity of the COVID-19 disease and the stress level in healthcare in terms of the number of hospitalizations, ICU occupancy and death.  Although hospitalizations and ICU occupancy relate directly to the healthcare stress level, all three can be seen as key indicators for the severity. Research has shown that vaccinations reduce the severity of the illness, which we clearly observe in our data too. Table \ref{tab:hospitalnumbers} shows the number of hospital and ICU beds, as well as the number of physicians and nurses in the five Nordic countries. According to these numbers, the healthcare system in Norway and Finland seem to be best positioned to handle a major health crisis such as a pandemic, although Finland has the lowest number of physicians. In countries with few hospital and ICU beds, it is even more important to reduce the stress level of the healthcare system. 

\begin{table}[h!]
    \centering 
    \caption{The number of hospital and ICU beds, doctors and nurses in the five Nordic countries. Source: https://tradingeconomics.com/}
    \label{tab:hospitalnumbers}
    \begin{tabular}{lcccc} \toprule
       Country& Hospital beds (per 1000)& ICU beds (per 100k)& Medical Doctors(per 1000) & Nurses(per 1000) \\ \midrule
        Denmark & 2.57&245.88&4.27&11.24 \\
         Finland&3.35&307.55&3.33&15.73\\
         Iceland&2.91&263.29&3.94&14.85\\
         Norway&3.60&342.52&5.53&20.83\\
         Sweden&2.22&234.50&4.32&11.49 \\ \bottomrule
    \end{tabular}
\end{table}

Coupled with that analysis, we also analyze the impact of the various vaccine types, vaccination rate on the spread of the virus in each age group for Denmark, Finland, Iceland, Norway and Sweden from the start of the vaccination period in December 2020 until the end of August 2021. Our findings from the analysis clearly shows that vaccines markedly reduce the number of new cases and the risk of serious illness. 

What can be clearly derived from our results is that the vaccination strategies vary to a large extent. The analyzed countries all have in common that they have led focused vaccination campaigns with high vaccination uptake early. However, there is a clear difference for Norway, which decided to select mRNA instead of a wide variety of vaccines for their citizens which resulted in a slower vaccination uptake compared to the other studied countries. Moreover, we see that Sweden has a steadier death rate and does therefore not have a similar decline in death rate early on as can be observed for the other studied countries. This may be a consequence of the different Sweden's strategy to battle the outbreak of the virus, which was different from the other countries \cite{yan2020countries}. In the Nordic countries, as well as in England, the oldest generations seem to be better protected against COVID-19 as seen by the low numbers of infections and less correlation with the total number of new infections. This can probably be attributed to the high ratio of fully vaccinated people among these cohorts. From the data, we see that Norway and Finland had different strategies from the other countries. Norway focused more on completing the vaccinations for one age group before moving on to a younger one, whereas Finland prioritized the administration of the first dose to as many as possible. The impact of these decisions is clear in the data. In Norway, infections among the younger populations have not yet reached diminishing numbers in 2021, which has been the case in the other countries. Currently the infection wave is growing fast for the $<15$ and $15-24$ age groups. In Finland however, the situation is graver. Not only are the infections numbers high and rising for $15-49$ year old people, but the number of hospitalizations, ICU occupancy and deaths, has not decreased to the same extent as in the other countries, which is concerning. Regarding the use of different vaccination types we show that the use of AstraZeneka was discontinued for most countries early. We also observe that Johnson \& Jonson / Jansen was primarily administered in Iceland but not used to the same extent in the other Nordic countries; Iceland later resulted in a booster dose for everyone that had accepted that particular vaccine type. Moreover, we show that Pfizer was used to the largest extent of all vaccine types in the data-set while Moderna has a steady spread across all countries. 

These results combined show that the wave structures and new infections is highly effected by the vaccination rate within and between various age groups in countries with high vaccination uptake. We also show that when the vaccination rate in a specific age group has reached a plateau, then the new infection rate follows and decreases drastically for that age group. Furthermore, we clearly show a wave on the rise for the youngest and least vaccinated part of the population now.

Regarding the data used, it is concerning that the data differs, depending on the country. For instance, there is no info on deaths divided by age for European countries whereas that data is readily available for England. Also, Norway did not report all data (ICU occupancy and vaccination type is missing), which has implications for all retrospective data analysis. The age groups in England are also significantly different compared to the ones used in the Nordic countries, making a comparison cross Europe as a whole difficult to achieve. Having unified data standards would make the work in cleaning the data, less time consuming. 

To sum up, we have examined the impact of vaccine uptake on the pandemic dynamics in the Nordic countries. In these countries, the vaccination campaigns has been notably efficient with a significantly high fraction of the population already vaccinated. Our analyses clearly demonstrate that a successful vaccination strategy is paramount to reduce stress on general health and its healthcare system. The results suggest that for countries where the fraction of unvaccinated people is still high, measures are required to either increase the vaccination rate or to continue with non-pharmaceutical measures such as social distancing.

Whether the resistance to vaccinations can be attributed to cultural factors such as obedience to authority and regulatory compliance remains an interesting open question. Early work on social order and disorder when it comes to non-pharmaceutical measures during COVID-19 has been published \cite{reicher2020order} whereas obedience to authority in relation to vaccinations has not been studied. We plan to address this and derive from the same set of data crucial information about human behavior on following government issued recommendations and regulations.  

\bibliography{references}

\section*{Author contributions statement}

 A.S.I, M.O., C.C., G.C. and F.S. gathered the data, analysed the results and wrote the paper. All authors reviewed the manuscript. 

\section*{Additional information}

\subsection*{eRG parameters tables}
\begin{center}
\begin{longtable}{ |p{1.5cm}||p{1.2cm}|p{2cm}|p{1.5cm}|p{1.5cm}|p{1.5cm}|p{2cm}|p{2cm}| }
\caption{eRG fit parameters.}\label{tab:eRG-Parameters} \\

\hline Country & Wave $\#$ & age group & $a$ & $\gamma$ & $t_0$ & $\%$ 1st dose & $\%$ 2nd dose\\ \hline
 \endfirsthead
 
\multicolumn{3}{c}%
{{\bfseries \tablename\ \thetable{} -- continued from previous page}} \\
\hline  Country & Wave $\#$ & age group & $a$ & $\gamma$ & $t_0$ & $\%$ 1st dose & $\%$ 2nd dose \\ \hline 
\endhead

\hline \multicolumn{3}{|r|}{{Continued on next page}} \\ \hline
\endfoot

\hline \hline
\endlastfoot

 Iceland & 1 & <15 & 171(2) & 1.51(7) & 3.2(4) & 0 & 0 \\
 & & 15-24 & 573(5) & 1.44(8) & 2.79(4) & 0 & 0\\
 & & 25-49 & 596(7) & 1.37(10) & 2.43(6) & 0 & 0\\
 & & 50-64 & 616(6) & 1.46(8) & 2.51(4) & 0 & 0\\
 & & 65-79 & 346(3) & 1.55(8) & 2.65(4) & 0 & 0\\
 & & 80+ & 188(6) & 1.7(4) & 3.0(2) & 0 & 0\\
 & & 50-79 & 513(5) & 1.48(8) & 2.54(4) & 0 & 0\\
 \hline
 & 2 & <15 & 601(8) & 0.80(4) & 7.7(7) & 0 & 0\\
 & & 15-24 & 1183(12) & 0.76(3) & 6.97(6) & 0 & 0\\
 & & 25-49 & 1172(20) & 0.63(3) & 7.33(10) & 0 & 0 \\
 & & 50-64 & 798(17) & 0.66(4) & 7.5(2) & 0 & 0\\
 & & 65-79 & 628(11) & 0.62(3) & 7.8(2) & 0 & 0\\
 & & 80+ & 750(33) & 1.2(3) & 8.4(2) & 0 & 0\\
 & & 50-79 & 733(14) & 0.64(4) & 7.6(2) & 0 & 0\\
 \hline
 & 3 & <15 & 169(7) & 0.65(7) & 7.4(3) & (18-24) 6.7 & 0.7\\
 & & 15-24 & 108(4) & 0.54(4) & 7.5(2) & (25-49) 7.1 & 1.9\\
 & & 25-49 & 189(7) & 0.48(4) & 6.9(3) & (50-59) 9.1 & 2.3\\
 & & 50-64 & 106(7) & 0.49(5) & 8.6(4) & (60-69) 18.2 & 5.1\\
 & & 65-79 & 32(2) & 0.6(1) & 5.8(4) & (70-79) 90.7 & 32.9 \\
 & & 80+ & 0 & 0 & 0 & (80+) 98.5 & 96.9 \\
 & & 50-79 & 77(4) & 0.48(5) & 8.2(3) & 30.3 & 10.0\\
 \hline 
 \hline
 Denmark & 1 & <15 & 1546(43) & 0.27(2) & 9.1(4) & 0 & 0\\
 & & 15-24 & 3454(170) & 0.26(2) & 11.8(5) & 0 & 0 \\
 & & 25-49 & 1991(65) & 0.25(2) & 10.2(4) & 0 & 0\\
 & & 50-64 & 1218(35) & 0.27(2) & 9.1(4) & 0 & 0\\
 & & 65-79 & 392(8) & 0.31(2) & 6.8(3) & 0 & 0\\
 & & 80+ & 225(4) & 0.32(3) & 4.9(3) & 0 & 0 \\
 & & 50-79 & 211(5) & 0.713(6) & 4.3(2) & 0 & 0\\
 \hline
 & 2 & <15 & 2989(99) & 0.31(2) & 17.9(3) & 0 & 0\\
 & & 15-24 & 5794(236) & 0.28(2) & 17.6(4) & 0 & 0\\
 & & 25-49 & 4394(173) & 0.29(2) & 18.2(4) & 0 & 0\\
 & & 50-64 & 3686(148) & 0.29(2) & 18.6(4) & 0 & 0\\
 & & 65-79 & 2017(108) & 0.28(2) & 19.3(5) & 0 & 0\\
 & & 80+ & 2874(189) & 0.32(3) & 20.9(5) & 0 & 0\\
 & & 50-79 & 2957(128) & 0.28(2) & 18.8(4) & 0 & 0\\
 \hline
 & 3 & <15 & 1546(44) & 0.27(2) & 9.1(4) & (18-24) 5.5 & 3.2\\
 & & 15-24 & 3454(170) & 0.26(2) & 11.8(5) & (25-49) 10.6 & 7.0\\
 & & 25-49 & 1991(65) & 0.25(2) & 10.2(4) & (50-59) 19.3 & 10.8\\
 & & 50-64 & 1218(35) & 0.27(2) & 9.1(4) & (60-69) 71.9 & 18.1\\
 & & 65-79 & 392(8) & 0.31(2) & 6.8(3) & (70-79) 97.1 & 76.8\\
 & & 80+ & 225(4) & 0.32(3) & 4.9(3) & (80+) 99.6 & 97.6\\
 & & 50-79 & 857(23) & 0.28(2) & 8.6(3) & 58.1 & 99.6\\
 \hline
 \hline
 Finland & 1 & <15 & 39.4(6) & 0.50(2) & 6.84(10) & 0 & 0 \\
 & & 15-24 & 130(3) & 0.58(4) & 5.8(2) & 0 & 0 \\
 & & 25-49 & 174(3) & 0.50(3) & 5.5(2) & 0 & 0 \\
 & & 50-64 & 148(3) & 0.57(3) & 4.9(1) & 0 & 0 \\
 & & 65-79 & 0 & 0 & 0 & 0 & 0 \\
 & & 80+ & 152(2) & 0.89(5) & 5.39(7) & 0 & 0 \\
 & & 50-79 & 106(2) & 0.59(4) & 4.8(2) & 0 & 0\\
 \hline
 & 2 & <15 & 457(9) & 0.321(7) & 16.1(2) & 0 & 0 \\
 & & 15-24 & 1128(63) & 0.27(2) & 14.3(6) & 0 & 0 \\
 & & 25-49 & 920(34) & 0.29(2) & 16.3(3) & 0 & 0 \\
 & & 50-64 & 572(25) & 0.29(2) & 16.7(4) & 0 & 0 \\
 & & 65-79 & 257(17) & 0.30(2) & 18.8(5) & 0 & 0 \\
 & & 80+ & 324(26) & 0.36(3) & 18.1(6) & 0 & 0 \\
 & & 50-79 & 428(21) & 0.29(2) & 17.0(4) & 0 & 0\\
 \hline
 & 3 & <15 & 1161(9) & 0.334(7) & 11.07(8) & (18-24) 2.1 & 0.6 \\
 & & 15-24 & 1704(14) & 0.308(8) & 9.97(10) & (25-49) 4.9 & 2.0 \\
 & & 25-49 & 1441(11) & 0.308(7) & 10.14(10) & (50-59) 9.5 & 2.3 \\
 & & 50-64 & 771(6) & 0.305(7) & 9.96(9) & (60-69) 14.3 & 20.2 \\
 & & 65-79 & 282(3) & 0.285(7) & 10.0(2) & (70-79) 20.2 & 0.8 \\
 & & 80+ & 310(4) & 0.248(8) & 9.3(2) & (80+) 78.3 & 6.5 \\
 & & 50-79 & 547(4) & 0.301(7) & 9.97(9) & 14.2 & 1.5\\
 \hline

 \hline
 Sweden & 1 & <15 & 78(11) & 0.29(4) & 16(2) & 0 & 0 \\
 & & 15-24 & 825(87) & 0.38(4) & 15.7(7) & 0 & 0 \\
 & & 25-49 & 1056(50) & 0.29(2) & 13.6(5) & 0 & 0 \\
 & & 50-64 & 1179(50) & 0.28(2) & 12.3(4) & 0 & 0 \\
 & & 65-79 & 576(21) & 0.30(3) & 9.1(4) & 0 & 0 \\
 & & 80+ & 1837(37) & 0.40(3) & 7.9(2) & 0 & 0 \\
 & & 50-79 & 910(39) & 0.27(2) & 11.4(5) & 0 & 0\\
 \hline
 & 2 & <15 & 1580(25) & 0.40(2) & 16.6(2) & 0 & 0 \\
 & & 15-24 & 7263(111) & 0.318(9) & 16.0(2) & 0 & 0 \\
 & & 25-49 & 7365(77) & 0.334(7) & 16.3(1) & 0 & 0 \\
 & & 50-64 & 6404(66) & 0.345(7) & 16.40(9) & 0 & 0 \\
 & & 65-79 & 2540(30) & 0.339(7) & 16.9(2) & 0 & 0 \\
 & & 80+ & 2975(42) & 0.365(8) & 16.73(9) & 0 & 0 \\
 & & 50-79 & 4663(50) & 0.343(7) & 16.5(1) & 0 & 0\\
 \hline
 & 3 & <15 & 3426(41) & 0.312(9) & 9.9(2) & (18-24) 4.5 & 1.7 \\
 & & 15-24 & 7913(82) & 0.314(8) & 9.7(2) & (25-49) 7.1 & 3.1 \\
 & & 25-49 & 7040(61) & 0.322(8) & 9.1(1) & (50-59) 10.5 & 5.2 \\
 & & 50-64 & 5658(53) & 0.328(9) & 8.9(1) & (60-69) 27.9 & 6.1 \\
 & & 65-79 & 1916(15) & 0.350(9) & 8.36(9) & (70-79) 79.0 & 12.8 \\
 & & 80+ & 1084(11) & 0.31(2) & 6.6(2) & (80+) 92.1 & 65.0 \\
 & & 50-79 & 3970(35) & 0.333(9) & 8.8(1) & 33.4 & 7.0\\
 \hline
 \hline
 Norway & 1 & <15 & 32(2) & 0.46(5) & 5.1(3) & 0 & 0 \\
 & & 15-24 & 137(3) & 0.63(5) & 3.7(2) & 0 & 0 \\
 & & 25-49 & 187(3) & 0.67(5) & 2.9(1) & 0 & 0 \\
 & & 50-64 & 196(3) & 0.79(6) & 2.58(9) & 0 & 0 \\
 & & 65-79 & 126(2) & 1.2(1) & 2.67(8) & 0 & 0 \\
 & & 80+ & 205(3) & 1.03(8) & 3.09(8) & 0 & 0 \\
 & & 50-79 & 166(3) & 0.90(7) & 2.61(9) & 0 & 0\\
 \hline
 & 2 & <15 & 703(31) & 0.33(3) & 11.6(4) & 0 & 0 \\
 & & 15-24 & 1636(94) & 0.28(3) & 11.5(6) & 0 & 0 \\
 & & 25-49 & 1335(66) & 0.28(3) & 11.6(5) & 0 & 0 \\
 & & 50-64 & 919(42) & 0.29(2) & 11.8(5) & 0 & 0 \\
 & & 65-79 & 432(14) & 0.31(2) & 11.5(3) & 0 & 0 \\
 & & 80+ & 546(22) & 0.31(2) & 12.5(3) & 0 & 0 \\
 & & 50-79 & 717(30) & 0.30(2) & 11.7(4) & 0 & 0\\
 \hline
 & 3 & <15 & 1415(21) & 0.36(2) & 9.4(2) & (18-24) 3.2 & 0.8 \\
 & & 15-24 & 2432(82) & 0.29(3) & 11.0(4) & (25-49) 4.6 & 1.7 \\
 & & 25-49 & 1478(20) & 0.34(2) & 9.3(2) & (50-59) 7.6 & 2.5 \\
 & & 50-64 & 903(10) & 0.34(2) & 9.2(2) & (60-69) 5.6 & 2.0 \\
 & & 65-79 & 311(2) & 0.43(2) & 8.58(8) & (70-79) 17.9 & 8.2 \\
 & & 80+ & 178(2) & 0.308(6) & 8.69(8) & (80+) 80.2 & 23.3 \\
 & & 50-79 & 657(7) & 0.36(2) & 9.0(2) & 9.5 & 3.8\\
 \hline

 \hline
 England & 1 & <4 & 28.3(4) & 0.51(3) & 6.0(1) & 0 & 0 \\
 & & <60 & 300(2) & 0.57(1) & 7.18(4) & 0 & 0 \\
 & & 5-9 & 38.9(6) & 0.425(7) & 10.63(8) & 0 & 0 \\
 & & 10-14 & 45.9(7) & 0.410(7) & 10.59(8) & 0 & 0 \\
 & & 15-19 & 101.7(6) & 0.510(5) & 9.06(3) & 0 & 0 \\
 & & 20-24 & 279(2) & 0.561(8) & 8.22(4) & 0 & 0 \\
 & & 25-29 & 401(3) & 0.58(1) & 7.43(4) & 0 & 0\\
 & & 30-34 & 397(3) & 0.58(1) & 7.30(4) & 0 & 0\\
 & & 35-39 & 366(2) & 0.58(1) & 7.21(4) & 0 & 0\\
 & & 40-44 & 412(3) & 0.59(1) & 6.99(4) & 0 & 0\\
 & & 45-49 & 467(3) & 0.60(2) & 6.83(4) & 0 & 0\\
 & & 50-54 & 491(3) & 0.59(1) & 6.78(3) & 0 & 0\\
 & & 55-59 & 478(3) & 0.58(2) & 6.71(4) & 0 & 0\\
 & & 60+ & 721(6) & 0.55(2) & 6.31(6) & 0 & 0\\
 & & 60-64 & 430(3) & 0.57(2) & 6.43(5) & 0 & 0\\
 & & 65-69 & 344(3) & 0.57(2) & 5.83(7) & 0 & 0\\
 & & 70-74 & 385(4) & 0.59(2) & 5.61(7) & 0 & 0\\
 & & 75-79 & 650(6) & 0.59(2) & 5.73(7) & 0 & 0\\
 & & 80-84 & 1148(10) & 0.56(2) & 6.14(6) & 0 & 0\\
 & & 85-89 & 1973(14) & 0.54(2) & 6.61(5) & 0 & 0\\
 & & 90+ & 3255(18) & 0.542(9) & 7.19(4) & 0 & 0\\
 \hline
 & 2 & <4 & 876(5) & 0.402(3) & 12.17(4) & 0 & 0 \\
 & & <60 & 2616(14) & 0.436(3) & 11.57(4) & 0 & 0 \\
 & & 5-9 & 1127(16) & 0.350(3) & 14.06(9) & 0 & 0 \\
 & & 10-14 & 2046(33) & 0.377(5) & 13.7(1) & 0 & 0 \\
 & & 15-19 & 3558(43) & 0.52(2) & 9.77(8) & 0 & 0 \\
 & & 20-24 & 3903(25) & 0.468(6) & 10.44(4) & 0 & 0 \\
 & & 25-29 & 3317(14) & 0.426(3) & 11.62(3) & 0 & 0\\
 & & 30-34 & 3103(11) & 0.434(2) & 11.81(2) & 0 & 0\\
 & & 35-39 & 2920(10) & 0.434(2) & 11.95(2) & 0 & 0\\
 & & 40-44 & 2906(8) & 0.440(2) & 12.07(2) & 0 & 0\\
 & & 45-49 & 2807(7) & 0.449(2) & 11.98(2) & 0 & 0\\
 & & 50-54 & 2701(8) & 0.464(2) & 11.82(2) & 0 & 0\\
 & & 55-59 & 2418(8) & 0.475(2) & 11.78(2) & 0 & 0\\
 & & 60+ & 1625(5) & 0.470(2) & 12.15(2) & 0 & 0\\
 & & 60-64 & 2006(6) & 0.481(2) & 11.82(2) & 0 & 0\\
 & & 65-69 & 1412(5) & 0.484(3) & 11.84(2) & 0 & 0\\
 & & 70-74 & 1137(4) & 0.491(2) & 11.96(2) & 0 & 0\\
 & & 75-79 & 1251(4) & 0.480(2) & 12.19(2) & 0 & 0\\
 & & 80-84 & 1614(5) & 0.457(2) & 12.57(2) & 0 & 0\\
 & & 85-89 & 2316(8) & 0.441(2) & 12.95(2) & 0 & 0\\
 & & 90+ & 3851(29) & 0.401(3) & 13.63(5) & 0 & 0\\
 \hline

 \hline
 England & 3 & <4 & 1643(9) & 0.550(9) & 5.64(4) & 0 & 0 \\
 & & <60 & 4213(18) & 0.63(1) & 5.18(3) & 0 & 0 \\
 & & 5-9 & 1624(9) & 0.55(1) & 5.02(4) & 0 & 0 \\
 & & 10-14 & 2410(14) & 0.58(2) & 4.36(4) & 0 & 0 \\
 & & 15-19 & 3771(15) & 0.630(9) & 4.83(3) & 0 & 0 \\
 & & 20-24 & 5344(21) & 0.68(1) & 5.22(3) & (18-24) 0.5 & 0.\\
 & & 25-29 & 5629(24) & 0.66(1) & 5.18(3) & 1.1 & 0.1\\
 & & 30-34 & 5578(26) & 0.63(1) & 5.28(4) & 1.2 & 0.1\\
 & & 35-39 & 5289(25) & 0.62(1) & 5.25(4) & 1.2 & 0.1\\
 & & 40-44 & 5167(23) & 0.62(1) & 5.15(3) & 1.5 & 0.1\\
 & & 45-49 & 4769(20) & 0.63(1) & 5.15(3) & 1.8 & 0.1\\
 & & 50-54 & 4578(18) & 0.638(9) & 5.28(3) & 2 & 0.1\\
 & & 55-59 & 4147(17) & 0.642(9) & 5.36(3) & 2.1 & 0.2\\
 & & 60+ & 2779(9) & 0.624(7) & 5.46(3) & - & -\\
 & & 60-64 & 3484(14) & 0.641(9) & 5.43(3) & 1.8 & 0.1\\
 & & 65-69 & 2348(10) & 0.646(9) & 5.38(3) & 0.9 & 0.1\\
 & & 70-74 & 1837(7) & 0.644(8) & 5.37(3) & 0.7 & 0.\\
 & & 75-79 & 2045(6) & 0.619(6) & 5.46(2) & 1.9 & 0.1\\
 & & 80-84 & 2654(7) & 0.600(5) & 5.48(2) & 25.2 & 4.3\\
 & & 85-89 & 4077(11) & 0.582(5) & 5.58(2) & 26.6 & 3.4\\
 & & 90+ & 6819(19) & 0.585(5) & 5.67(2) & 22.9 & 3.7\\
 \hline
 & 4 & <4 & 1011(12) & 0.570(6) & 8.48(5) & 0 & 0 \\
 & & <60 & 3580(59) & 0.610(9) & 8.48(5) & 0 & 0 \\
 & & 5-9 & 2099(18) & 0.652(6) & 7.83(4) & 0 & 0 \\
 & & 10-14 & 4481(56) & 0.629(8) & 8.02(5) & 0 & 0 \\
 & & 15-19 & 6543(80) & 0.620(9) & 8.02(5) & 0 & 0 \\
 & & 20-24 & 6673(61) & 0.642(7) & 7.57(4) & (18-24) 61.8 & 20.9\\
 & & 25-29 & 5452(137) & 0.62(2) & 8.3(1) & 60.9 & 27.1\\
 & & 30-34 & 4493(156) & 0.60(2) & 8.7(2) & 64.1 & 40.6\\
 & & 35-39 & 3529(109) & 0.60(2) & 8.6(2) & 69.2 & 52.3\\
 & & 40-44 & 3059(71) & 0.60(2) & 8.49(9) & 75.4 & 66.6\\
 & & 45-49 & 2456(53) & 0.59(1) & 8.49(9) & 81.3 & 75.1\\
 & & 50-54 & 2086(59) & 0.58(2) & 8.7(2) & 86 & 82.4\\
 & & 55-59 & 1659(53) & 0.58(2) & 8.9(2) & 88.5 & 85.3\\
 & & 60+ & 818(31) & 0.56(2) & 9.4(2) & - & -\\
 & & 60-64 & 1277(51) & 0.57(2) & 9.2(2) & 90.4 & 87.7\\
 & & 65-69 & 898(36) & 0.58(2) & 9.4(2) & 92.3 & 90.7\\
 & & 70-74 & 627(20) & 0.59(2) & 9.2(2) & 94.5 & 93.3\\
 & & 75-89 & 566(16) & 0.569(8) & 9.4(1) & 95.5 & 94.4\\
 & & 80-84 & 584(21) & 0.510(8) & 10.0(2) & 95.6 & 93.9\\
 & & 85-89 & 686(33) & 0.468(7) & 10.7(2) & 95.6 & 93.7\\
 & & 90+ & 663(23) & 0.495(7) & 10.2(2) & 94 & 91.8\\

\end{longtable}
\end{center}


\end{document}